\documentclass[preprint,3p]{elsarticle}

\usepackage{amssymb}
\usepackage{color}

\journal{Journal of Computational Physics}

\begin{document}
\newcommand{\HALF}{\frac{1}{2}}
\newcommand{\pd}[2]{\frac{\partial #1}{\partial #2}}
\newcommand{\DS}{\displaystyle}
\renewcommand{\vec}[1]{\mathbf{#1}}
\newcommand{\tens}[1]{\mathsf{#1}}
\newcommand{\red}[1]{\color{red} #1 \color{black}}
\newcommand{\blue}[1]{\color{blue} #1 \color{black}}

\begin{frontmatter}



\title{A Second-Order Unsplit Godunov Scheme for Cell-Centered MHD:
       the CTU-GLM scheme.}


\author[ut]{Andrea Mignone\corref{cor1}}
\ead{mignone@ph.unito.it}

\author[ut]{Petros Tzeferacos}
\ead{petros.tzeferacos@to.infn.it}

\cortext[cor1]{Corresponding Author}

\address[ut]{Dipartimento di Fisica Generale, Universit\'a degli studi di Torino}

\begin{abstract}
We assess the validity of a single step Godunov scheme for 
the solution of the magneto-hydrodynamics equations in more 
than one dimension. 
The scheme is second-order accurate and the temporal 
discretization is based on the dimensionally unsplit 
Corner Transport Upwind (CTU) method of Colella.
The proposed scheme employs a cell-centered representation
of the primary fluid variables (including magnetic field) and
conserves mass, momentum, magnetic induction and energy.
A variant of the scheme, which breaks momentum and energy conservation,
is also considered.
Divergence errors are transported out of the domain 
and damped using the mixed hyperbolic/parabolic
divergence cleaning technique by Dedner et al. 
(J. Comput. Phys., 175, 2002).
The strength and accuracy of the scheme are verified by 
a direct comparison with the eight-wave formulation (also employing
a cell-centered representation) and with the popular constrained 
transport method, where magnetic field components retain a staggered
collocation inside the computational cell.
Results obtained from two- and three-dimensional test problems 
indicate that the newly proposed scheme is robust, accurate 
and competitive with recent implementations of the constrained 
transport method while being considerably easier to implement in 
existing hydro codes.


\end{abstract}


\end{frontmatter}


\section{Introduction}
\label{sec:intro}
%
%
%

A primary aspect in building stable and robust Godunov type schemes
for the numerical solution of the compressible magnetohydrodynamics (MHD) 
equations relies on an accurate way to control the 
solenoidal property of the magnetic field while preserving 
the conservation properties of the underlying physical laws.
Failure to fulfill either requisite has been reported as a potential 
hassle leading to unphysical effects such as plasma
acceleration in the direction of the field, incorrect jump conditions,
wrong propagation speed of discontinuities and odd-even decoupling, see \cite{Toth00,BK04}.
A comprehensive body of literature has been dedicated to this subject
and several strategies to enforce the $\nabla\cdot\vec{B}=0$ condition
in Godunov-type codes have been proposed, see for example
\cite{ZMC94, RJF95, RMJF98, BS99, Toth00} and, more recently,
\cite{Balsara04, LdZ04, GS05, Rossmanith06, MBMR08}. 
The robustness of one method over another can be established on a practical base 
by extensive numerical testing, see \cite{Toth00, BK04}.

In a first class of schemes, the magnetic field is discretized 
as a cell-centered quantity and the usual formalism already 
developed for the Euler equation can be extended in a natural way. 
Cell-centered methods are appealing since the extensions to 
adaptive and/or unstructured grids are of straightforward implementation. 
Moreover, the same interpolation scheme and stencil 
used for the other hydrodynamic variables can be easily adapted since
all quantities are discretized at the same spatial location,
thus facilitating the extension to schemes possessing higher than 
second order accuracy.
Unfortunately, numerical methods based on a cell-centered discretization 
do not naturally preserve Gauss's
law of electromagnetism, even if $\nabla\cdot\vec{B}=0$ initially.
In the approach suggested by Powell \cite{Powell94, Powell99}, Gauss's law for
magnetism is discarded in the derivation of the MHD equations and the resulting 
system of hyperbolic laws is no longer conservative by the appearance
of a source term proportional to $\nabla\cdot\vec{B}$.
Although the source term should be physically zero at the continuous level, 
Powell showed that its inclusion changes the character of the equations by 
introducing an additional eighth wave corresponding to the propagation of 
jumps in the component of magnetic field normal to a given interface.
A different approach is followed in the projection scheme 
\cite{BB80, ZMC94, RJF95, Crockett05}, where a Helmholtz-Hodge 
decomposition is applied to resolve $\vec{B}$ as the sum of an 
irrotational and a solenoidal part, associated with scalar 
and vector potentials.
A cleaning step allows to recover the divergence-free magnetic
field by subtracting the unphysical contribution coming
from the irrotational component at the extra cost
of solving a Poisson equation. 
In the approach of Dedner et al. \cite{Dedner02}, 
the divergence-free constraint is enforced 
by solving a modified system of conservation laws where
the induction equation is coupled to a generalized
Lagrange multiplier. Dedner et al. showed that the choice of 
mixed hyperbolic/parabolic correction offers both propagation and
dissipation of divergence errors with the maximal admissible
characteristic speed, independently of the fluid velocity.
This approach preserves the full conservation form of the original
MHD system at the minimal cost of introducing one additional
variable in the system and will be our scheme of choice.
Finally, Torrilhon \cite{Torrilhon05} (see also \cite{AT08}) showed 
a general procedure to modify the inter-cell fluxes in the framework of 
a flux distribution scheme that preserves the value of a certain 
discrete divergence operator in each control volume.


A different strategy is followed in the constrained transport (CT)
methods, originally devised by \cite{EH88} and later built into 
the framework of shock-capturing Godunov methods by a number of 
investigators, e.g., \cite{BS99, Balsara04, LdZ04, GS05, GS08}.
In this class of schemes, the magnetic field has a staggered 
representation whereby the different components live on the face they 
are normal to.
Hydrodynamic variables (density, velocity and pressure) retains 
their usual collocation at the cell center. 
CT schemes preserve the divergence-free condition to machine accuracy 
in an integral sense since the magnetic field is treated as a surface 
averaged quantity and thus more naturally updated using Stokes' theorem.
This evolutionary step involves the construction of a line-averaged electric 
field along the face edges, thereby requiring some sort of reconstruction 
or averaging of the electromotive force from the face center 
(where different components are usually available as face 
centered upwind Godunov fluxes) to the edges.
A variety of different strategies have been suggested, including 
simple arithmetic averaging \cite{BS99, RMJF98}, solution of 2-D Riemann problems
\cite{LdZ04, FHT06} or other somewhat more empirical approaches 
\cite{GS05,GS08,LD09}.
The staggered collocation of magnetic and electric field variables 
in CT schemes makes their extension to adaptive grids rather
arduous and costly. Besides, significant efforts have to be spent in
order to develop schemes with spatial accuracy of order higher than second.
An alternative constrained transport method, based on the direct solution 
of the magnetic potential equation (thus avoiding staggered grids), 
has been presented by \cite{Rossmanith06}. 
 
In the present work we propose a new fully unsplit Godunov scheme for 
multidimensional MHD, based on a combination 
of the Corner Transport Upwind of \cite{Colella90} and the 
mixed hyperbolic/parabolic divergence cleaning technique of 
\cite{Dedner02} (CTU-GLM).
The proposed scheme has second order accuracy in both space and time
and adopts a cell-centered spatial collocation (no staggered mesh) 
of all flow variables, including the magnetic field. 
The scheme is fully conservative in mass, momentum, magnetic induction and energy
and the divergence-free constraint is enforced via a mixed hyperbolic/parabolic
correction which avoids the computational cost associated with an 
elliptic cleaning deriving from a Hodge projection.
A variant of the scheme, which introduces divergence source terms breaking
the conservative properties of some equations, is also presented.
We assess the accuracy and robustness of the scheme by a 
direct quantitative comparison with the 8-wave formulation of \cite{Powell99} 
and the recently developed constrained transport method of \cite{GS05, GS08}.
Other similar implementations may be found in \cite{FHT06,LD09}. 
The comparison is conveniently handled using the PLUTO code for computational 
astrophysics \cite{Mignone07} where both cell-centered and staggered-mesh 
implementations are available.

Our motivating efforts are driven by issues of simplicity, efficiency and 
flexibility. In this sense, the benefits offered by a method where all of 
the primary flow variables are discretized at the same spatial location 
considerably ease the extension to adaptive grids, to more complex physics
and to schemes with higher than second order accuracy. 
The latter possibility will be explored in a companion paper.


\section{The Constrained GLM-MHD Equations}
\label{sec:eqns}
%
%
%

In the approach of \cite{Dedner02}, the divergence constraint of
the magnetic field (Gauss's law) is coupled to Faraday's equation 
by introducing a new scalar field function or generalized Lagrangian 
multiplier $\psi$.
The second and third Maxwell's equations are thus replaced by 
\begin{equation}\label{eq:mod_maxwell}
 \left\{\begin{array}{rcl}
\DS  \nabla\cdot\vec{B} &= & 0 \,,\\ \noalign{\medskip}
\DS  \pd{\vec{B}}{t}    &= & \nabla\times\left(\vec{v}\times\vec{B}\right) \,,
 \end{array}\right. 
 \qquad \Longrightarrow \qquad 
 \left\{\begin{array}{rcl}
\DS  {\cal D}(\psi) + \nabla\cdot\vec{B} &=& 0 \,, \\ \noalign{\medskip}
\DS  \pd{\vec{B}}{t} + \nabla\psi &=& \nabla\times\left(\vec{v}\times\vec{B}\right) \,,
  \end{array}\right.
\end{equation}
where ${\cal D}$ is a linear differential operator.
Dedner et al. built this approach into the MHD equations and 
showed that a satisfactory explicit approximation 
may be obtained by choosing a mixed hyperbolic/parabolic
correction, according to which ${\cal D}(\psi) = c_h^{-2}\partial_t\psi + 
c_p^{-2}\psi$ where $c_h$ and $c_p$ are constants.
Direct manipulation of the modified Maxwell's equations
(\ref{eq:mod_maxwell}) leads to the telegraph equation,
\begin{equation}
\label{eq:telegraph}
 \frac{\partial^2\psi}{\partial t^2} + \frac{c_h^2}{c_p^2}\pd{\psi}{t} = 
  c_h^2\Delta\psi \,,
\end{equation}
which implies that divergence errors are propagated to the 
domain boundaries at finite speed $c_h$ and
decay with time and distance. The constant ratio $c_h^2/c_p^2$,
which has the dimension of inverse time, sets the damping rate. 
In the limiting case of $c_p\to\infty$, one retrieves the simple
hyperbolic correction and Eq. (\ref{eq:telegraph}) reduces to 
an ordinary wave equation.

The GLM-Maxwell's equations (\ref{eq:mod_maxwell}) can be coupled to the 
equations of magnetohydrodynamics written in their conservative form.
The resulting system is called the generalized Lagrange multiplier (GLM) 
formulation of the MHD equations (GLM-MHD) and is comprised of the 
following nine evolution equations:
\begin{equation}
\label{eq:glm_mhd}
\begin{array}{rcl}
\DS \pd{\rho}{t} + \nabla\cdot\left(\rho\vec{v}\right) &=& 0 \,, \\ \noalign{\medskip}
\DS \pd{\vec{(\rho\vec{v})}}{t} + \nabla\cdot\left[
   \rho\vec{v}\vec{v}^T 
 -     \vec{B}\vec{B}^T 
 + \tens{I}\left(p + \frac{\vec{B}^2}{2}\right)\right] &=& 0\,,\\ \noalign{\medskip}
\DS \pd{\vec{B}}{t} + \nabla\cdot\left(\vec{v}\vec{B}^T - \vec{B}\vec{v}^T\right) 
 + \nabla\psi    &=& 0\,,\\ \noalign{\medskip}
\DS \pd{E}{t} + \nabla\cdot\left[
  \left(E + p + \frac{\vec{B}^2}{2}\right)\vec{v} - 
  \left(\vec{v}\cdot\vec{B}\right)\vec{B}\right]& =& 0\,,\\ \noalign{\medskip}
\DS \pd{\psi}{t} + c_h^2 \nabla\cdot\vec{B} & = & \DS -\frac{c_h^2}{c_p^2}\psi \,,
\end{array}
\end{equation}
where $\rho$, $\vec{v}$, $p$  and $\vec{B}$ are the mass density, 
velocity, gas pressure and magnetic field, respectively.
Total energy $E$ and gas pressure are related by the ideal gas law, 
$E = p/(\Gamma-1) + \rho\vec{v}^2/2 + \vec{B}^2/2$, where $\Gamma$ is
the specific heat ratio.
Notice that we have conveniently switched, using vector identities,
to the divergence form of the induction equation,
more appropriate for the cell-centered finite volume formalism. 
The constrained GLM-MHD equations (\ref{eq:glm_mhd})
are hyperbolic and fully conservative in all flow variables with the 
exception of the unphysical scalar field $\psi$ which satisfies a 
non-homogeneous equation with a source term. 
Divergence errors propagate with speed $c_h$ independently 
of the flow velocity, thus avoiding accumulation in 
presence of stagnation points. 
The presence of the source term is responsible for damping divergence errors 
as they propagate.

Dedner et al. also considered a slightly different constrained formulation,
in which the Lorentz force term in the MHD equations is directly 
derived from the GLM-Maxwell equations. In this case, the system 
(\ref{eq:glm_mhd}) is extended by an additional source term 
on the right hand side, namely
\begin{equation}
\label{eq:seglm}
 \vec{S}_{EGLM} = \left[0, -(\nabla\cdot\vec{B})\vec{B}, 
                        \vec{0}, -\vec{B}\cdot\nabla\psi,0\right]^T\,,
\end{equation}
where the non-zero entries correspond to the momentum and energy
equations. Dedner called the system (\ref{eq:glm_mhd}) augmented with 
the source term (\ref{eq:seglm}) on its right hand side 
the \emph{extended} GLM (EGLM) formulation of the MHD equations.
Although the system breaks conservation of energy and momentum, it still
holds some attractive features and we found it, in our experience, 
a more robust scheme in presence of strong discontinuity propagating
through highly magnetized environments.

\section{The CTU-GLM scheme}
\label{sec:numscheme}
%
%
%
%
%

We now illustrate the detailed steps of our new 
cell-centered numerical scheme. The derivation
is shown for the conservative GLM scheme, 
whereas modifications relevant to the EGLM formulation 
are described in \S\ref{sec:eglm}.

We adopt a Cartesian system of coordinates and re-write 
the system of equations in (\ref{eq:glm_mhd}) as
\begin{equation}\label{eq:system}
\pd{}{t} \left(\begin{array}{c}
 \rho     \\ \noalign{\medskip}
 \rho v_d \\ \noalign{\medskip}
 B_d \\ \noalign{\medskip}
 E        \\ \noalign{\medskip}
 \psi \end{array}\right)
 +
 \sum_{l=x,y,z}
 \pd{}{l}\left(\begin{array}{c}
 \rho v_l  \\ \noalign{\medskip}
 \rho v_dv_l - B_dB_l + \delta_{dl}\left(p + \vec{B}^2/2\right)  \\ \noalign{\medskip}
  B_dv_l - B_lv_d + \delta_{dl}\psi \\ \noalign{\medskip}
 \left(E + p + \vec{B}^2/2\right)v_l - \left(\vec{v}\cdot\vec{B}\right)B_l    \\ \noalign{\medskip}
 c_h^2 B_l \end{array}\right)
  = 
\left(\begin{array}{c}
  0 \\ \noalign{\medskip}
  0 \\ \noalign{\medskip}
  0 \\ \noalign{\medskip}
  0 \\ \noalign{\medskip}
 -c_h^2/c_p^2\psi  \end{array}\right) \,,
\end{equation}
where $d,l = x,y,z$ label the different component and flux contributions
in the three directions while $\delta_{dl}$ is the delta Kronecker symbol.
The system of equations given in (\ref{eq:system}) is advanced in time
by solving the homogeneous part separately from the source term 
contribution, in an operator-split fashion:
\begin{equation}
  \vec{U}^{n+1} = {\cal S}^{\Delta t/2}
                  {\cal A}^{\Delta t}
                  {\cal S}^{\Delta t/2}\vec{U}^n
\end{equation}
where ${\cal A}$ and ${\cal S}$ are the advection and source step 
operators separately described in \S\ref{sec:advection} and 
\S\ref{sec:source}, respectively.

\subsection{Advection Step}
\label{sec:advection}
%
%
%
%

During the homogeneous step, we adopt a numerical discretization 
of (\ref{eq:system}) based on the corner transport upwind (CTU) 
method of \cite{Colella90}. 
For simplicity, we will assume hereafter an equally-spaced grid 
with computational cells centered in $(x_i,y_j,z_k)$ having size 
$\Delta x\times\Delta y\times\Delta z$.
For the sake of exposition, we omit the subscript $(i,j,k)$ from cell 
centered quantities while keeping the half increment 
index notation when referring to the interfaces, e.g., 
$\rho_{j+\HALF} \equiv \rho_{i,j+\HALF,k}$.
An explicit second order accurate discretization of Eqns. (\ref{eq:system}), 
based on a time-centered flux evaluation, reads
\begin{equation}\label{eq:update}
 \vec{U}^{n+1} = \vec{U}^n - \Delta t^n\left[
   \frac{\vec{F}^{n+\HALF}_{i+\HALF} - \vec{F}^{n+\HALF}_{i-\HALF}}{\Delta x}
 + \frac{\vec{G}^{n+\HALF}_{j+\HALF} - \vec{G}^{n+\HALF}_{j-\HALF}}{\Delta y}
 + \frac{\vec{H}^{n+\HALF}_{k+\HALF} - \vec{H}^{n+\HALF}_{k-\HALF}}{\Delta z}
\right]  \,,
\end{equation}
where $\vec{U} = (\rho, \rho\vec{v}, \vec{B}, E, \psi)$ is the state vector
of conservative variables.
The expression in square brackets provides a conservative discretization of 
the divergence operator appearing in the original conservation laws 
with $\vec{F}$, $\vec{G}$ and $\vec{H}$ being suitable numerical approximations 
to the flux contributions in (\ref{eq:system}) 
coming from the $l=x,y,z$ directions, respectively. 
In the CTU approach, numerical fluxes are computed by solving a Riemann 
problem between suitable time-centered left and right states, i.e.,
\begin{equation}\label{eq:ctu_flux}
\vec{F}^{n+\HALF}_{i+\HALF} = {\cal R}\left(\vec{V}^{n+\HALF}_{i,+}, 
                                            \vec{V}^{n+\HALF}_{i+1,-}\right)
\,,\quad
\vec{G}^{n+\HALF}_{i+\HALF} = {\cal R}\left(\vec{V}^{n+\HALF}_{j,+}, 
                                            \vec{V}^{n+\HALF}_{j+1,-}\right)
\,,\quad
\vec{H}^{n+\HALF}_{i+\HALF} = {\cal R}\left(\vec{V}^{n+\HALF}_{k,+}, 
                                            \vec{V}^{n+\HALF}_{k+1,-}\right)\,,
\end{equation}
where $\vec{V}=(\rho,\vec{v},\vec{B}, p,\psi)^T$ is the state vector
of primitive variables and ${\cal R}(\cdot,\cdot)$ denotes the
flux obtained by means of a Riemann solver, see \S\ref{sec:riem}.
The corner-coupled states, $\vec{V}^{n+\HALF}_{i,+}$ and $\vec{V}^{n+\HALF}_{i+1,-}$, 
are computed via a Taylor expansion consisting of 
an evolutionary step in the direction normal to a given interface
(\S\ref{sec:normal}) followed by a correction step involving 
transverse flux gradients (\S\ref{sec:transverse}).
The algorithm requires a total of $6$ solution to the Riemann problem per zone 
per step.

The time increment $\Delta t^n$ is computed via the Courant-Friedrichs-Levy 
(CFL) condition:
\begin{equation}\label{eq:deltat}
 \Delta t^n = C_a\frac{\min\left(\Delta x, \Delta y, \Delta z\right)}
                      {\max_{i,j,k}\left(|v_x| + c_{f,x}, |v_y| + c_{f,y}, |v_z| + c_{f,z}\right)}\,,
\end{equation}
where the maximum and minimum are taken over all zones and $c_{f,x}, c_{f,y}, c_{f,z}$ 
are the fast magneto-sonic speeds in the three directions, see \S\ref{sec:normal}. 
$C_a$ is the Courant number and, for the 6-solve CTU presented here,
is restricted to $C_a < 1$ in two dimensions and $C_a< 1/2$ in three dimensions.

\subsubsection{Normal Predictors}
\label{sec:normal}
%
%

During the computation of the normal predictors, we take advantage of 
the primitive (or quasi-linear) form of the equations. 
By discarding contributions from $y$ and $z$ and considering the 
reconstruction process in the $x$ direction only, one has
\begin{equation}\label{eq:prim}
 \pd{\vec{V}}{t} + \tens{A}_x\pd{\vec{V}}{x} = 
  \vec{S}_{B_x}\pd{B_x}{x} + \vec{S}_\psi\pd{\psi}{x} \,,
\end{equation}
where the $9\times 9$ matrix 
\begin{equation}
 \tens{A}_x = \left(\begin{array}{ccccccccc}
 v_x & \rho     &  0   & 0    & 0     &    0      &    0      &    0   & 0 \\ \noalign{\medskip}
  0  & v_x      &  0   & 0    & 0     &  B_y/\rho &  B_z/\rho  & 1/\rho & 0 \\ \noalign{\medskip}
  0  &   0      & v_x  & 0    & 0     & -B_x/\rho &    0      &    0   & 0 \\ \noalign{\medskip}
  0  &   0      &  0   & v_x  & 0     &   0       & -B_x/\rho &    0   & 0 \\ \noalign{\medskip}
  0  &   0      &  0   & 0    & 0     &   0       &    0      &    0   & 1 \\ \noalign{\medskip}
  0  & B_y      & -B_x & 0    & 0     &   v_x     &    0      &    0   & 0 \\ \noalign{\medskip}
  0  & B_z      &  0   & -B_x & 0     &    0      &   v_x     &    0   & 0 \\ \noalign{\medskip}
  0  & \Gamma p &  0   & 0    & 0     &    0      &    0      &   v_x  & 0  \\ \noalign{\medskip}
  0  &   0      &  0   & 0    & c_h^2 &    0      &    0      &    0   & 0
\end{array}\right)\,, 
\end{equation}
is the usual matrix of the MHD equations in primitive form plus the addition of a 
fifth row and a ninth column. The source terms $\vec{S}_{B_x}$ and
$\vec{S}_\psi$ are of crucial importance for the accuracy of the scheme 
in multi-dimensions \cite{Crockett05, GS05, LD09} and take the form
\begin{equation}\label{eq:prim_src}
\vec{S}_{B_x} = 
\DS \left[0,\frac{B_x}{\rho}, \frac{B_y}{\rho}, \frac{B_z}{\rho},
             0, v_y, v_z, 
        -(\Gamma - 1)\vec{v}\cdot\vec{B}, 0\right]^T \,,\quad
\vec{S}_{\psi}  = 
\DS   \Big[0, 0 , 0, 0, 0, (\Gamma-1)B_x,0\Big]^T \,.
\end{equation}
The matrix $\tens{A}_x$ of the quasi-linear form is diagonalizable 
with the same eigenvalues as the ordinary MHD equations 
plus two new additional entries $c_h$ and $-c_h$, for a total
of $9$ characteristic waves:
\begin{equation}
 \lambda^{1,9} = \mp c_h     \,,\quad
 \lambda^{2,8} = v_x \mp c_f \,,\quad
 \lambda^{3,7} = v_x \mp c_a \,,\quad
 \lambda^{4,6} = v_x \mp c_s \,,\quad
 \lambda^5     = v_x \,,
\end{equation}
where
\begin{equation}\label{eq:eigenvalues}
c_{f,s} = \sqrt{\frac{1}{2\rho}\left(\Gamma p + |\vec{B}|^2 \pm 
          \sqrt{\left(\Gamma p + |\vec{B}|^2\right)^2 
                - 4\Gamma p B_x^2}\,\right)}
\,,\quad
c_a = \frac{\left|B_x\right|}{\sqrt{\rho}} \,,
\end{equation}
are the fast magneto-sonic ($c_f$ with the $+$ sign), 
slow magneto-sonic ($c_s$ with the $-$ sign) and Alfv\'en velocities.       
The two additional modes $\pm c_h$ are decoupled from the remaining ones 
and corresponds to waves carrying jumps in $B_x$ and $\psi$.
The constant $c_h$ gives the speed of propagation of local 
divergence errors and is chosen to be the maximum speed compatible
with the time step restriction, in other words 
\begin{equation}\label{eq:ch}
 c_h = \max_{i,j,k}\left(|v_x| + c_{f,x}, |v_y| + c_{f,y}, |v_z| + c_{f,z}\right)\,,
\end{equation}
Finally, the corresponding left ($\vec{l}^k$) and right ($\vec{r}^k$) eigenvectors are 
given in Appendix \ref{app:eigenv}.

Using the characteristic decomposition of the quasi-linear form (\ref{eq:prim}), 
we extrapolate $\vec{V}(x_i,t^n)$ from the cell center to the edges $x_{i\pm\HALF}$ for 
a time increment $\Delta t^n/2$. 
During this step we only consider the contribution
of those waves traveling from the center to the given interface and 
discard any interaction between neighbor cells.
The resulting construction yields the normal predictors 
\begin{equation}\label{eq:norm_pred}
  \vec{V}^{*}_{i,\pm} = \vec{V}^n_i + \frac{1}{2}\sum_{k:\lambda^k_i \gtrless 0} 
  \left(\pm 1 - \frac{\lambda^k_i \Delta t^n}{\Delta x}\right)\Delta\vec{V}^k_i  
  + \frac{\Delta t^n}{2\Delta x}\left(\vec{S}^n_{B_x,i}\Delta B_x + \vec{S}^n_{\psi,i}\Delta\psi\right)\,,
\end{equation}
where only positive waves ($\lambda^k_i > 0$, $k=1,...,9$) 
contribute to the left of the $i+\HALF$ interface ($i,+$) while only 
negative waves ($\lambda^k_i < 0$) are considered to the right of
the $i-\HALF$ interface ($i,-$).
The undivided differences $\Delta B_x$ and $\Delta\psi$ 
may be computed using a standard centered finite difference
approximation.
The jump contribution from the $k-$th characteristic field is denoted 
with $\Delta\vec{V}^k_i=\Delta w^k_i \vec{r}^k_i$ where 
$\vec{r}^k_i$ is the corresponding right eigenvector and 
$\Delta w^k_i$ is a limited slope in the $k-$th characteristic variable, 
\begin{equation}
 \Delta w^k_i = {\rm Lim}\left(\vec{l}_i^k\cdot\Delta\vec{V}^n_{i+\HALF} ,
                              \vec{l}_i^k\cdot\Delta\vec{V}^n_{i-\HALF}\right) \,,
\end{equation}    
where $\Delta\vec{V}^n_{i\pm\HALF}=\pm\left(\vec{V}^n_{i\pm 1} - \vec{V}^n_i\right)$,
$\vec{l}^k_i$ is the $k-$th primitive left eigenvector
and ${\rm Lim}(\cdot,\cdot)$ is a limiter function, e.g. 
\begin{equation}\label{eq:limiter}
  {\rm Lim}(\delta_-,\delta_+) = 
  \frac{\rm{sign}(\delta_-) + \rm{sign}(\delta_+)}{2}
                  \min\left(\beta|\delta_-|, \beta|\delta_+|, 
  \frac{\delta_-+\delta_+}{2}\right) \,.
\end{equation}
Usually taking $\beta = 2$ gives the largest compression. 
However, for problems involving strong shocks, 
we found setting $\beta=1$ for nonlinear fields (fast and slow shocks)
and $\beta=2$ for the linear fields to give a more robust recipe.

\subsubsection{Transverse Predictors}
\label{sec:transverse}
%
%

Once the normal predictor states have been computed, we solve a 
Riemann problem at constant $y-$ and $z-$ faces 
to obtain the transverse fluxes, e.g.,
\begin{equation}\label{eq:transv_flux}
 \vec{G}^{*}_{j+\HALF} =
 {\cal R}\left(\vec{V}^{*}_{j,+}, \vec{V}^{*}_{j+1,-}\right)
\,, \quad
 \vec{H}^{*}_{k+\HALF} =
 {\cal R}\left(\vec{V}^{*}_{k,+}, \vec{V}^{*}_{k+1,-}\right)
\,,
\end{equation}
where left and right states have been computed during the normal
predictor stages in the $y$ and $z$ direction.
The solution of the Riemann problem follows the guidelines illustrated 
in \S\ref{sec:riem}, where the linear sub-system formed by the longitudinal
magnetic field component and the Lagrange multiplier is preliminary solved
before a standard $7-$wave Riemann solver is applied.
Transverse flux gradients are then added to the normal predictors 
(\ref{eq:norm_pred}) once they are transformed back to conservative 
variables. This yields the corner coupled states:
\begin{equation}\label{eq:cc_states}
 \vec{U}^{n+\HALF}_{i\pm\HALF} = \vec{U}^{*}_{i\pm\HALF}  
  - \frac{\Delta t}{2}\left(
    \frac{\vec{G}^{*}_{j+\HALF} - \vec{G}^{*}_{j-\HALF}}{\Delta y}
 +  \frac{\vec{H}^{*}_{k+\HALF} - \vec{H}^{*}_{k-\HALF}}{\Delta z}
  \right)   \,,
\end{equation}
where $\vec{U}^*$ is obtained by converting $\vec{V}^*$ to conservative
variables. 

We recall that the starting point in the derivation of Eq. 
(\ref{eq:cc_states}) may be viewed, in its simplest form, as a 
first order Taylor expansion around the cell center $(x_i,t^n)$,
\begin{equation}
 \vec{U}^{n+\HALF}_{i\pm\HALF} \approx \vec{U}^{n}_{i} \pm 
                                    \pd{\vec{U}^n_i}{x}\frac{\Delta x}{2}
                                  + \pd{\vec{U}^n_i}{t}\frac{\Delta t}{2}
  \approx \left(\vec{U}^n_i \pm \pd{\vec{U}^n_i}{x}\frac{\Delta x}{2}
          - \frac{\Delta t}{2}\pd{\vec{F}^n_i}{x}\right)
          - \frac{\Delta t}{2}\left(\pd{\vec{G}^*_i}{y} +\pd{\vec{H}^*_i}{z}\right) \,,
\end{equation}
where the temporal derivative $\partial\vec{U}/\partial{t}$ has been
replaced, in the second expression, by taking advantage of the original 
conservation law and the different terms have been grouped according to 
the step in which they are computed (i.e., Eq \ref{eq:norm_pred} and 
Eq \ref{eq:cc_states}).
In this perspective, the input states entering in the computation of 
the transverse fluxes (\ref{eq:transv_flux}) may be slightly modified 
by $O(\Delta t^2)$ in order to more accurately represent the 
$\nabla\cdot\vec{B}$ term in the construction of the scalar multiplier $\psi$.
To better understand this minor correction, we rewrite the $\psi$ 
component of the interface states (\ref{eq:cc_states}) in 2D using, for 
the sake of simplicity, a simple MUSCL-Hancock step during the normal predictor:
\begin{equation}\label{eq:psicc}
 \psi^{n+\HALF}_{\pm} = \psi^n\pm \frac{\Delta\psi^n}{2} - 
                        \frac{c_h^2\Delta t}{2}\left[\frac{\Delta B^n_x}{\Delta x} 
     +\frac{B^*_{y,j+\HALF} - B^*_{y,j-\HALF}}{\Delta y}\right] \,.
\end{equation}
Clearly, the multidimensional terms approximating $\nabla\cdot\vec{B}$ 
in the square bracket of Eq. (\ref{eq:psicc}) split into a normal 
($\Delta B^n_x$) and a transverse ($B^*_{y,j+\HALF}-B^*_{y,j-\HALF}$)
directional contribution.
Since the first one is taken at time level $n$ while the second
term comes from solving a Riemann problem between normal 
predictors in the $y$ direction (extrapolated a $t^n+\Delta t^n/2$), 
these contributions are not taken at the same time level but are 
spaced by $\Delta t^n/2$.
In practice, from the tests included here and several others we 
found evidence that a better balance is achieved if one replaces, 
in the input states of (\ref{eq:transv_flux}), the longitudinal 
field component with its interpolated value at 
time level $n$, i.e., $B^*_{y,j,\pm} \to B^n_{y} \pm \Delta B^n_{y}/2$ 
(or, equivalently with the value obtained by setting 
$\Delta t = 0$ in Eq. \ref{eq:norm_pred}).
Note that this is a second-order correction that does not alter
the accuracy of the scheme and only affects the solution of
the Riemann problem in computing the transverse fluxes 
(\ref{eq:transv_flux}) but does not concern 
the definitions of the normal predictors.
Although this is not an essential step, it was found to improve 
the accuracy in the numerical tests presented in \S\ref{sec:numtest}.
%
 
\subsection{Solving the Riemann Problem}
\label{sec:riem}
%
%
%
%

In the case of the GLM-MHD equations, left and right input states to 
the Riemann solver ${\cal R}(\cdot, \cdot)$ bring a set of $9$ jumps 
propagating along the $7$ standard characteristic MHD waves 
(i.e. fast, slow, rotational pairs and one entropy modes) as well as 
$2$ additional modes carrying jumps only in the normal (longitudinal) 
component of $\vec{B}$ and $\psi$.
Nonetheless, when solving a one-dimensional Riemann problem at a zone interface
(say the $x$ direction), these additional waves are decoupled from the 
remaining ones and are described by the  
$2\times 2$ linear hyperbolic system
\begin{equation}\label{eq:2x2}
\left\{\begin{array}{rcl}
\DS \pd{B_x}{t}  & = & \DS -\pd{\psi}{x}       \\ \noalign{\medskip}
\DS \pd{\psi}{t} & = & \DS -c_h^2\pd{B_x}{x}  \,.
\end{array}\right.
\end{equation}
For a generic pair of left and right input states $(B_{x,L},\psi_L)$ and 
$(B_{x,R}, \psi_R)$, the Godunov flux of the system (\ref{eq:2x2}) 
can be computed exactly as
\begin{equation}\label{eq:bstar}
 B^*_x = \frac{B_{x,L} + B_{x,R}}{2}   - \frac{1}{2c_h}\left(\psi_R-\psi_L\right)
 \,,\quad
 \psi^*  = \frac{\psi_{L} + \psi_{R}}{2} - \frac{c_h}{2} \left(B_{x,R} - B_{x,L}\right)
 \,.
\end{equation}
This allows to carry out the solution of the $2\times 2$ linear Riemann 
problem separately before using any standard $7$-wave Riemann solver
for the one-dimensional MHD equations. 
The longitudinal component of the magnetic field $B^*_x$,
preliminary computed with (\ref{eq:bstar}), enters hence the ordinary 
Riemann flux computation as a constant parameter.

In other words, given the arbitrary left and right states 
$\vec{V}_L$ and $\vec{V}_R$, input to the Riemann problem, we compute 
\begin{equation} 
  {\cal R}\left(\vec{V}_L, \vec{V}_R\right) = {\cal R}_7\left(\vec{V}^*_L, \vec{V}^*_R\right)
%
\end{equation}
where $\vec{V}^*_S$ ($S=L,R$) is the same as $\vec{V}_S$ with 
$\left(B_{x,S},\psi_S\right)$ replaced by $\left(B_x^*, \psi^*\right)$ 
and ${\cal R}_7$ is a standard $7-$wave Riemann solver.
In this work, we will employ the linearized Riemann solver of Roe, 
in the version of \cite{CG97}.

\subsection{Source Step}
\label{sec:source}
%
%
%
%

During the source step we solve the initial value
problem given by the last of equations (\ref{eq:glm_mhd}) 
without the $\nabla\cdot\vec{B}$ term, that is,
\begin{equation}\label{eq:psi_source}
  \pd{\psi}{t} = -\frac{c_h^2}{c_p^2}\psi\,,
\end{equation}  
supplemented with the initial 
condition $\psi^{(0)}$ given by the output of the 
most recent step.
The constant $c^2_p$ has the dimension of length squared over time and thus can 
be regarded as a diffusion coefficient. 
Dedner et al. prescribe an optimal value $c_p^2/c_h=0.18$ 
independently of the mesh spacing; however, we suspect this definition to be
incomplete, since $c_p^2/c_h$ has the dimension of length and 
thus it is \emph{not} a dimensionless quantity.
Our numerical experiments indicate that divergence errors
are minimized when the parameter $\alpha = \Delta hc_h/c_p^2$ (where 
$\Delta h = \min(\Delta x, \Delta y,\Delta z)$) lies in the range
$\alpha\in[0,1]$, depending on the particular problem.
In first approximation this value can be regarded as grid-independent although 
we have verified a weak tendency to decrease as the mesh thickens.
Using the definition of $\alpha$, Eq. (\ref{eq:psi_source}) can be integrated 
exactly for a time increment $\Delta t^n$, yielding
\begin{equation} \label{eq:source_exp}
 \psi^{(\Delta t^n)} = \psi^{(0)}\exp\left(-\alpha\frac{c_h}{\Delta h/\Delta t}\right)\,,
 \quad \mathrm{with} \quad
  \alpha = \Delta h\,\frac{c_h}{c_p^2}\,.
\end{equation}
Note that, when $c_h$ is chosen using Eq. (\ref{eq:ch}), the
argument of the exponential becomes simply $(-C_a\alpha)$.
Finally, we comment out that the dimensionless $\alpha$ parameter can be 
regarded as the ratio of the diffusive and advective time scales, i.e.,
$\alpha=\Delta t_d/\Delta t_a$, where $\Delta t_d = \Delta h^2/c_p^2$ 
and $\Delta t_a = \Delta h/c_h$.

\subsection{Modifications for the Extended GLM (EGLM) formulation}
\label{sec:eglm}
%
%
%
%

The extended GLM-MHD (EGLM-MHD) equations may be derived from 
the primitive MHD equations rather than the conservative ones,
\cite{Dedner02}.
In this approach, the divergence part of the Lorentz force is 
added to the momentum flux and an additional source term,
given by (\ref{eq:seglm}), is introduced into the system.
The construction of the normal predictor states carried out in
\S\ref{sec:normal} remains the same with the exception of 
the source terms (\ref{eq:prim_src}) which must be replaced by
\begin{equation}\label{eq:eglm_prim_src}
\vec{S}_{B_x} = 
\DS \left[0, 0, 0, 0,
             0, v_y, v_z, 
        -\left(\Gamma-1\right)\vec{v}\cdot\vec{B}, 0\right]^T \,,\quad
\vec{S}_{\psi}  = \vec{0}\,.
\end{equation}
Since the corner coupled states in Eq. (\ref{eq:cc_states}) are obtained
in conservative variables, they must also be augmented with the source term 
contribution (Eq. \ref{eq:seglm}) and thus replaced by
\begin{equation}\label{eq:eglm_cc_states}
 \vec{U}^{n+\HALF}_{i\pm\HALF} \rightarrow 
 \vec{U}^{n+\HALF}_{i\pm\HALF} + \frac{\Delta t}{2}\left(\vec{S}^n_{EGLM,y}+\vec{S}^n_{EGLM,z}\right)  \,.
\end{equation}
Likewise, the final update Eq. (\ref{eq:update}) becomes
\begin{equation}\label{eq:eglm_update}
 \vec{U}^{n+1} \rightarrow \vec{U}^{n+1} + \Delta t\left(
       \vec{S}^{n+\HALF}_{EGLM,x} + \vec{S}^{n+\HALF}_{EGLM,y} + \vec{S}^{n+\HALF}_{EGLM,z}\right)\,.
\end{equation}
In Eq. (\ref{eq:eglm_cc_states}) and (\ref{eq:eglm_update}) 
we have split the source term into contributions 
coming from the derivatives in the $x$, $y$ and $z$ directions.
For each term we take advantage of the upwind fluxes computed 
in the corresponding direction during the Riemann solver step. 
For example, during the $y-$sweep we compute the momentum 
and energy sources in $\vec{S}_{EGLM,y}$ as
\begin{equation}
  -\vec{B}\pd{B_y}{y} \approx -\vec{B}\left(\frac{B^*_{y,j+\HALF}-B^*_{y,j-\HALF}}{\Delta y}\right)
    \,,\quad
  -B_y\pd{\psi}{y} \approx -B_y\left(\frac{\psi^*_{j+\HALF}-\psi^*_{j-\HALF}}{\Delta y}\right)\,,
\end{equation}
where $B^*_y$ and $\psi^*$ follows from the solution of the linear 
$2\times 2$ Riemann problem (\ref{eq:bstar}).
The cell-centered magnetic field is evaluated at $t^n$ for the
computation of the corner coupled states (\ref{eq:eglm_cc_states}) 
and by averaging to cell-center the final interface values
for the final update, Eq. (\ref{eq:eglm_update}).

\section{Numerical Tests}
\label{sec:numtest}
%
%
%

We now proceed to a direct verification of the CTU-GLM and CTU-EGLM algorithms 
developed in the previous sections.
A test suite of standard two- and three-dimensional MHD problems has been 
selected in order to monitor and quantify the accuracy of the proposed schemes.
For the sake of comparison, we extend the verification process to other
two well known methods, namely, Powell's eight wave formulation
\cite{Powell99} based on a cell-centered approach and the constrained 
transport (CT) scheme of \cite{GS05, GS08} using a staggered formulation.
The four selected algorithms, ``GLM'', ``EGLM'', ``8W'' and ``CT'', 
have been built into the CTU methodology and have been implemented 
in the current distribution of the PLUTO code for astrophysical 
gas-dynamics \cite{Mignone07} 
available at \emph{http://plutocode.to.astro.it}. 
Adopting the same numerical framework provides a practical way for
a convenient and extensive inter-scheme comparison.

In the following test problems the scalar field function $\psi$ will be 
always initialized to zero and thus omitted from the definition of
the initial conditions.
Moreover, unless otherwise stated, the specific heat ratio will be set 
to $\Gamma = 5/3$ and the default Courant number is set to $C_a = 0.8$ 
in two dimensions and $C_a=0.4$ in three dimensions.
Errors for any flow quantity $Q$ are computed using the $L_1$ discrete 
norm defined by
\begin{equation}\label{eq:L1}
 \epsilon_1(Q) = \frac{1}{N_xN_yN_z}\sum_{i,j,k}\left|Q_{i,j,k} - Q_{i,j,k}^{\rm ref}\right| 
\end{equation}
where $N_x$, $N_y$ and $N_z$ are the number of points in the three 
directions, $Q_{i,j,k}^{\rm ref}$ is a reference solution
and the summation extends to all grid zones.

\subsection{Propagation of Circularly polarized Alfv\'en Waves}
\label{sec:alfv}
%
%

Circularly polarized Alfv\'en waves are an exact nonlinear solution of the 
compressible MHD equations thus providing an excellent code benchmark.
For a planar wave propagating along the $x$ direction with angular frequency 
$\omega$ and wave number $k$, the transverse components of velocity and magnetic 
fields trace circles in the $yz$ plane and the solution can be written as 
\begin{equation}\label{eq:alfv_1D}
  \left(\begin{array}{c}
  v_x  \\ \noalign{\medskip}
  v_y  \\ \noalign{\medskip}
  v_z  
\end{array}\right)
=
 \left( \begin{array}{c}
  v_{0x} \\ \noalign{\medskip}
  v_{0y} + A\sin\phi \\ \noalign{\medskip}
  v_{0z} + A\cos\phi 
\end{array}\right)
\,,\quad
  \left(\begin{array}{c}
  B_x  \\ \noalign{\medskip}
  B_y  \\ \noalign{\medskip}
  B_z  
\end{array}\right)
=
 \left( \begin{array}{c}
  c_a\sqrt{\rho} \\ \noalign{\medskip}
  \mp\sqrt{\rho}A\sin\phi \\ \noalign{\medskip}
  \mp\sqrt{\rho}A\cos\phi 
\end{array}\right)\,,
\end{equation}
where $\phi = kx - \omega t$, $\omega/k = v_{0x}\pm c_a$ is the corresponding 
phase velocity ($c_a=1$ is the Alfv\'en speed) and $A=1/10$ is the wave amplitude. 
The  plus or minus sign corresponds to right or left propagating waves, 
respectively.
The constants $v_{0x}, v_{0y}, v_{0z}$ give the translational velocity components 
in the three directions.
Density and pressure remain constant and equal to their 
initial values $\rho_0=1$ and $p_0=0.1$ since torsional Alfv\'en waves 
do not involve any compression.

Here we consider a rotated version of the one-dimensional 
solution given by (\ref{eq:alfv_1D}) and specify the orientation 
of the wave vector $\vec{k}=(k_x,k_y,k_z)$ in a three dimensional 
space $x,y,z$ through the angles $\alpha$ and $\beta$ such that
\begin{equation}
  \tan\alpha = \frac{k_y}{k_x} \,,\quad
  \tan\beta  = \frac{k_z}{k_x} \,.
\end{equation}
The full 3D solution is then recovered by rotating the original 
one dimensional frame by an angle $\gamma=\tan^{-1}(\cos\alpha\tan\beta)$ 
around the $y$ axis and subsequently by an angle $\alpha$ around 
the $z$ axis. The resulting transformation leaves scalar quantities 
invariant and produce vectors rotation 
$\vec{q}\to \tens{R}_{\gamma\alpha}\vec{q}$, where
\begin{equation}\label{eq:rot_mat}
 \tens{R}_{\gamma\alpha} =
 \left(\begin{array}{ccc}
  \cos\alpha\cos\gamma &  -\sin\alpha & -\cos\alpha\sin\gamma  \\ \noalign{\medskip}
  \sin\alpha\cos\gamma &   \cos\alpha & -\sin\alpha\sin\gamma  \\ \noalign{\medskip}
  \sin\gamma           &   0          & \cos\gamma
 \end{array}\right)\,,\quad
 \tens{R}^{-1}_{\gamma\alpha} = 
 \left(\begin{array}{ccc}
  \cos\alpha\cos\gamma &   \sin\alpha\cos\gamma & \sin\gamma  \\ \noalign{\medskip}
  -\sin\alpha          &   \cos\alpha           &  0          \\ \noalign{\medskip}
 -\cos\alpha\sin\gamma & -\sin\alpha\sin\gamma  & \cos\gamma   
 \end{array}\right)\,,
\end{equation}
are the rotation matrix and its inverse whereas $\vec{q}$ is a three-dimensional
vector.
Note that $\phi$ is now given by $\phi = \vec{k}\cdot\vec{x} - \omega t$  
where $\omega = |\vec{k}|(v_{0x} \pm c_a)$ and
\begin{equation}
 \left|\vec{k}\right| = k_x\sqrt{1 + \tan^2\alpha + \tan^2\beta}
\end{equation}
is the wavenumber corresponding to a wavelength 
$\lambda = 2\pi/|\vec{k}|$ and period $T = 2\pi/\omega$.

In order to ensure correct periodicity we assume, without loss of generality,
$k_x = 2\pi$ and pattern the computational domain such that one wave period is 
prescribed in each grid direction, i.e., $x\in [0,1]$, $y\in[0,1/\tan\alpha]$ and
$z\in[0,1/\tan\beta]$.
Also, for the tests discussed here, we consider standing waves and thus
set $v_{0x} = v_{0y} = v_{0z} = 0$.
With these definitions the wave returns into the original position 
after one period $T = \lambda/c_a$ with 
\begin{equation}\label{eq:alfv_period}
 T = \frac{1}{\sqrt{1 + \tan^2\alpha + \tan^2\beta}}\,.
\end{equation}
%

\subsubsection{Two-Dimensional Propagation}
%
%
%
%

\begin{table}[!ht]
\caption{Errors (in $L_1$ norm) and orders of accuracy for the two and 
         three-dimensional circularly polarized Alfv\'en wave tests.
         The first and second columns refer to the numerical scheme and 
         the number of points in the $x$ direction.
         Columns 3-4 and 5-6 show the result obtained in the 2D problem 
         with Courant number of $C_a=0.8$ and $C_a=0.4$, respectively.
         The last two columns corresponds to the three dimensional case.}
\label{tab:cpa}
\centering
\begin{tabular*}{\textwidth}{@{\extracolsep{\fill}} lr rrrrrr}\hline
        &   &  \multicolumn{2}{c}{2D, $C_a=0.8$}  &   \multicolumn{2}{c}{2D, $C_a=0.4$}  & \multicolumn{2}{c}{3D, $C_a=0.4$} \\
  \cline{3-4} \cline{5-6} \cline{7-8}
Scheme & $N_x$  & $L_1$ Error  & $L_1$ order & $L_1$ Error  & $L_1$ order       
                & $L_1$ Error  & $L_1$ order \\ 
 \noalign{\smallskip} \hline\noalign{\smallskip}
\hline
GLM
 &  16 &      2.46E-002 &    -   &      2.60E-002 &     -  &      3.19E-002 &     -  \\ \noalign{\smallskip}
 &  32 &      4.56E-003 &   2.43 &      5.17E-003 &   2.33 &      5.66E-003 &   2.50 \\ \noalign{\smallskip}
 &  64 &      1.16E-003 &   1.97 &      1.27E-003 &   2.03 &      1.15E-003 &   2.30 \\ \noalign{\smallskip}
 & 128 &      3.19E-004 &   1.87 &      3.02E-004 &   2.07 &      3.03E-004 &   1.92 \\ \noalign{\smallskip}
 & 256 &      8.48E-005 &   1.91 &      7.01E-005 &   2.11 &      8.05E-005 &   1.91 \\ \noalign{\smallskip}
\hline CT
 &  16 &      2.54E-002 &    -   &      2.79E-002 &     -  &      3.44E-002 &     -  \\ \noalign{\smallskip}
 &  32 &      4.96E-003 &   2.36 &      7.09E-003 &   1.98 &      5.57E-003 &   2.63 \\ \noalign{\smallskip}
 &  64 &      1.16E-003 &   2.09 &      1.90E-003 &   1.90 &      1.18E-003 &   2.24 \\ \noalign{\smallskip}
 & 128 &      2.76E-004 &   2.08 &      4.25E-004 &   2.16 &      3.24E-004 &   1.86 \\ \noalign{\smallskip}
 & 256 &      6.73E-005 &   2.04 &      9.32E-005 &   2.19 &      9.67E-005 &   1.75 \\ \noalign{\smallskip}
\hline 8W
 &  16 &      2.60E-002 &     -  &      2.81E-002 &     -  &      3.37E-002 &     -  \\ \noalign{\smallskip}
 &  32 &      5.19E-003 &   2.32 &      7.28E-003 &   1.95 &      5.44E-003 &   2.63 \\ \noalign{\smallskip}
 &  64 &      1.22E-003 &   2.09 &      1.88E-003 &   1.95 &      1.37E-003 &   1.99 \\ \noalign{\smallskip}
 & 128 &      2.96E-004 &   2.05 &      4.02E-004 &   2.22 &      3.45E-004 &   1.99 \\ \noalign{\smallskip}
 & 256 &      7.29E-005 &   2.02 &      8.40E-005 &   2.26 &      8.79E-005 &   1.97 \\ \noalign{\smallskip}
\hline
\end{tabular*}
\end{table}

\begin{figure}[!h]\centering
\includegraphics[width=\textwidth]{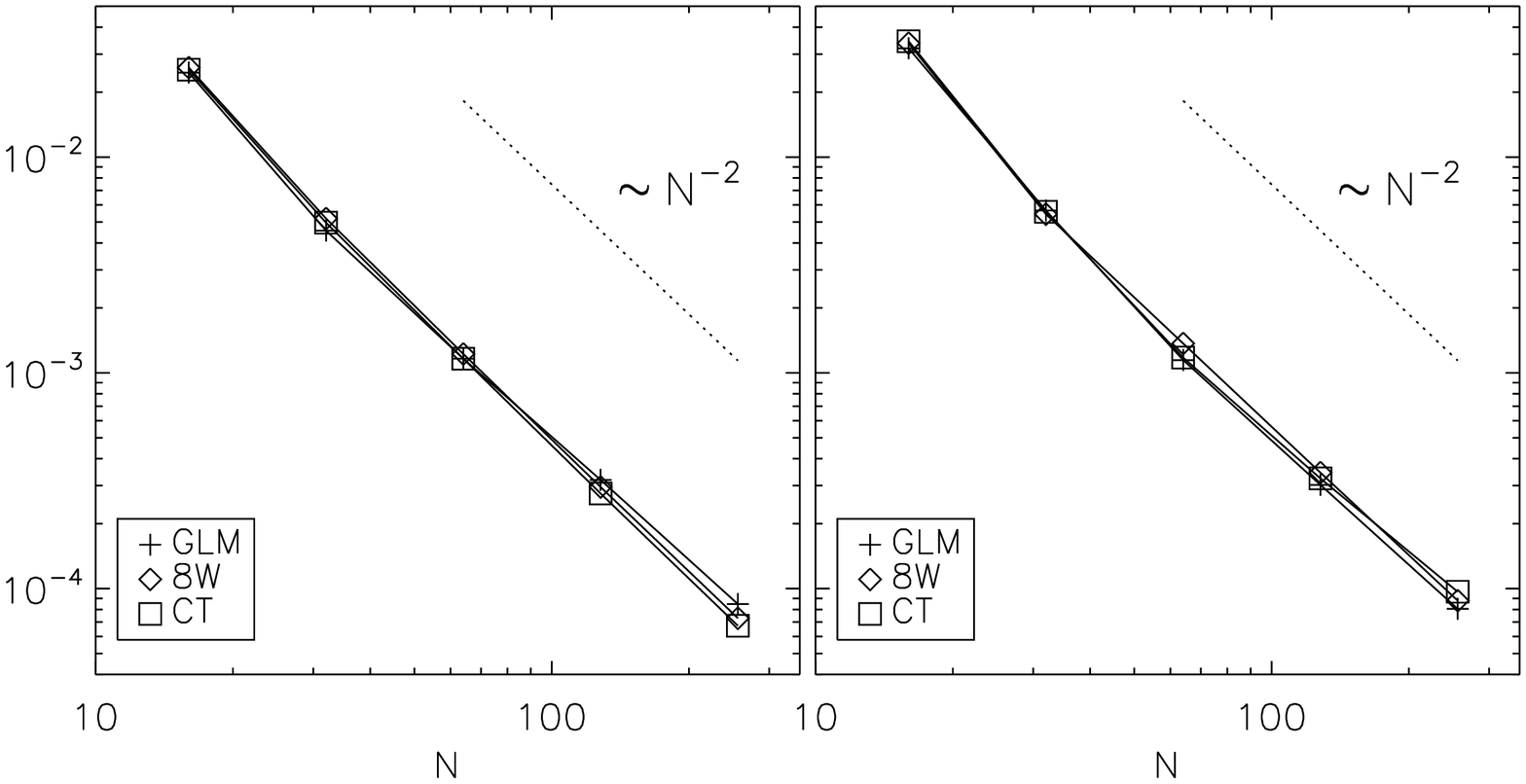}
 \caption{$L_1$ norm errors for the 2D (left) and 3D (right) circularly 
          polarized Alfv\'en wave test problem. 
          Each symbol refers to results obtained with the GLM (plus sign), 
          CT (square) and Powell's eight wave (rhombus) methods, while
          the dotted line gives the ideal second-order convergence slope.
          The Courant number $C_a=0.8$ and the final time step 
          is $1/\sqrt{5}$ (left) and $1/3$ right.}
 \label{fig:cpa}
\end{figure}

We begin by considering two dimensional propagation choosing
$\tan\alpha=2$, $\beta = 0$ in accordance with \cite{LdZ04,GS05,LD09}.
Computations are carried out for exactly one wave period 
($t=T=1/\sqrt{5}$) on the computational box $[0,1]\times[0,1/2]$ 
with $N_x\times N_y$ points, where $N_y=N_x/2$. 
Errors, computed as 
$\sqrt{\epsilon_1(B_x)^2+\epsilon_1(B_y)^2+\epsilon_1(B_z)^2}$,
are reported in Table \ref{tab:cpa} and plotted as function of 
the mesh size, $N_x=16,...,256$, in the left panel of Fig \ref{fig:cpa}.

Selected schemes (CT, GLM and 8W) produce comparable errors 
and show essentially second-order accuracy.
We notice that decreasing the Courant number to $C_a=0.4$ has the 
effect of slightly reducing the errors for GLM 
at large resolution but not for CT and 8W.
From Table \ref{tab:cpa}, in fact, one can see that, when $N_x=256$, 
the error is reduced from $\sim 8.5\cdot 10^{-5}$ to $\sim 7\cdot 10^{-5}$
for GLM, while it grows from $6.7\cdot 10^{-5}$ to $9.3\cdot 10^{-5}$ 
for the CT scheme.

We have found that the solution is very weakly dependent on the 
$\alpha$ parameter and the errors are minimized when $\alpha=0$.
Besides, we repeated the computations with the EGLM formulation and 
observed essentially the same level of accuracy with no 
particular improvement over GLM.

\subsubsection{Three-Dimensional Propagation}
%
%
%
%

In three dimensions we follow \cite{GS08} and set 
$\tan\alpha = \tan\beta = 2$ so that the resulting computational 
box is given by $x\in[0,1]$, $y,z\in[0,1/2]$ discretized on 
$N_x\times N_x/2 \times N_x/2$ grid points.
Computations are followed for one wave period ($T=1/3$) and repeated, 
with $C_a=0.4$, on increasingly finer grids corresponding to 
$N_x=16,32,64,128,256$.
The right panel in Figure \ref{fig:cpa} shows that all schemes
meet the expected order of accuracy providing comparable errors, 
as found in the two-dimensional case.
A more quantitative comparison can be made by inspecting 
the last two columns of Table \ref{tab:cpa}, where 
one can see that GLM performs slightly better than the other 
schemes.

\subsection{Nonlinear smooth flow}
%

\begin{figure}[!h]
\centering
\includegraphics[width=0.7\textwidth]{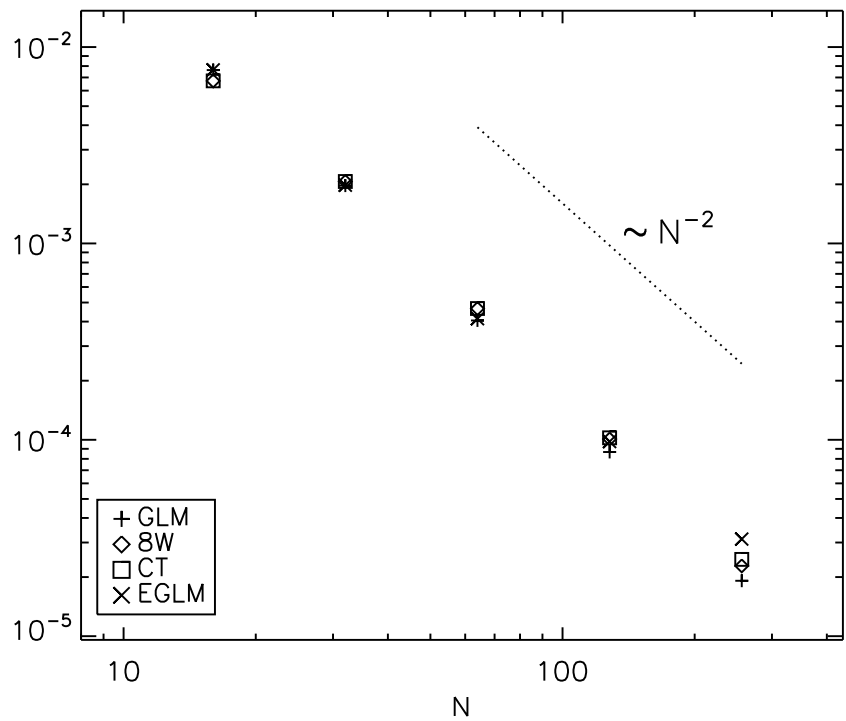}
 \caption{$L_1$ norm errors for the non-linear, smooth flow
          test problem at $t=0.2$. 
          The different symbols refer to computations carried out with the 
          GLM (plus signs), EGLM (ex signs), CT (squares) and Powell's
          eight wave method (rhombus) with Courant number
          $C_a=0.8$. The dotted line gives the ideal second-order convergence 
          slope.}
 \label{fig:nonlin}
\end{figure}

In the next example we consider the evolution of a fully 
nonlinear smooth flow where, unlike the previous example, 
all waves (linear and nonlinear) are triggered.
Following \cite{Torrilhon05,AT08} we specify a periodic computational box in 
Cartesian coordinates, spanning from $-1$ to $1$ in the $x$ and $y$ directions
with initial conditions given by
\begin{equation}\begin{array}{lcl}
 \rho &=& \DS  \frac{3}{2} + \HALF \sin(\pi x) + 
                      \frac{1}{4} \cos(\pi y) \,, \\ \noalign{\medskip}
 (v_x,\,v_y) &=& \DS \left[1 + \frac{1}{2} \sin (\pi y) +
                               \frac{1}{4} \cos (\pi x),\,
                           1 + \frac{1}{4} \sin (\pi x) + 
                               \frac{1}{2} \cos (\pi y)\right] \,,\\ \noalign{\medskip} 
 (B_x,\,B_y) &=& \DS \left(\HALF,\,1\right) \,,
\end{array}
\end{equation}
where $p = 1/4$, while $v_z = B_z = 0$.
Integration terminates at $t=0.2$, before the formation of any discontinuous 
feature. 
A resolution study is carried out for all schemes and compared to a reference 
solution obtained on $2048^2$ zones with the CT scheme. 
The error, shown in Fig. (\ref{fig:nonlin}) as a function of the number of cells,  
is computed as a quadratic mean of the $L_1$ norm errors (given by Eq. \ref{eq:L1})
of the primitive variables.
All schemes are second-order accurate with comparable errors, with the GLM 
approach giving slightly better results than the others at the largest resolution
($256$ zones).

\subsection{Shock Tube Problems}
%
%

One dimensional shock tubes have proven to be valuable benchmarks 
in order to assess the ability of the scheme to capture both 
continuous and discontinuous flow features.
The rotated multidimensional versions considered in the following 
may be used to check the strength of the numerical method 
in preserving the original planar symmetry through an 
oblique propagation.

\subsubsection{Two-dimensional shock tube}
%
%

In the first shock tube, taken from \cite{Toth00}, we consider an initial 
discontinuity with left and right states given by 
$(\rho,v_1,v_2,B_1,B_2,p)_L = (1,10,0,5/\sqrt{4\pi},5/\sqrt{4\pi},20)$ 
and  
$(\rho,v_1,v_2,B_1,B_2,p)_R =  (1,-10,0,5/\sqrt{4\pi},5/\sqrt{4\pi},1)$ 
respectively. The subscripts ``1'' and ``2'' give the directions 
perpendicular and parallel to the initial discontinuity.
The initial condition is then rotated on a Cartesian grid $(x,y)$
using the transformation defined by Eq. (\ref{eq:rot_mat}) with 
$\alpha = \tan^{-1}2$ and $\beta=\gamma=0$. 

Since the magnetic field is initially uniform, $\nabla \cdot {\bf B} = 0$ is 
trivially ensured at $t=0$.
The computational domain spans from 0 to 1 in the $x$ direction and from 
0 to $2/N_x$ in the $y$ direction with $N_x\times 2$ computational zones.
Outflow boundaries are set at the rightmost and leftmost sides of the 
box whereas for any flow variables $q$ at the upper and lower boundaries we 
impose the translational invariance $q(i,j) = q(i\pm\delta i, j\pm\delta j)$ where
$(\delta i, \delta j) = (2,-1)$ with the plus (minus) sign holding 
at the upper (lower) boundary. 
Computations terminate before the fast shocks reach the boundaries, 
at $t=0.08\cos\alpha$. 

\begin{figure}[!h]
\centering
\includegraphics[width=\textwidth]{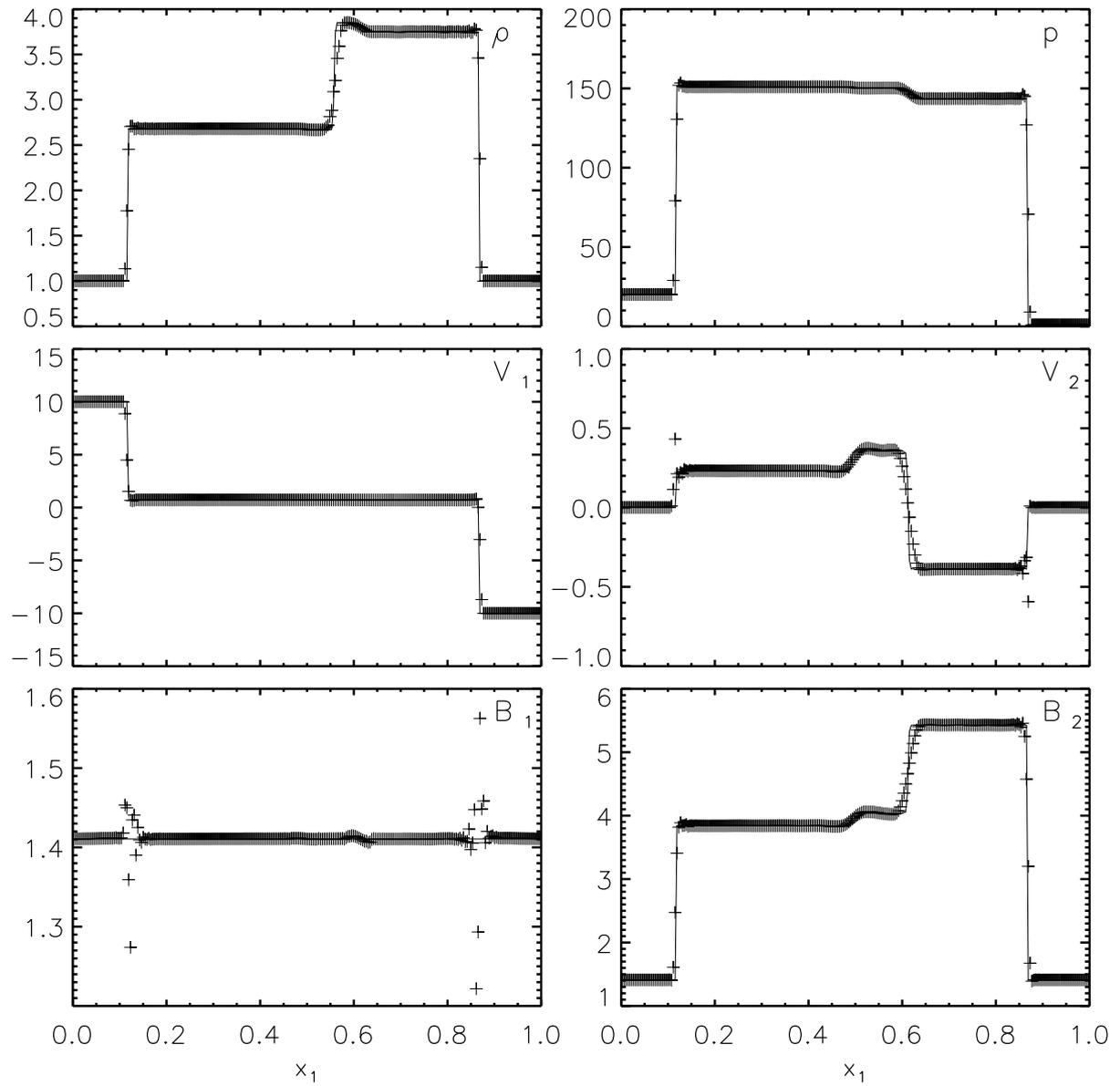}
 \caption{Primitive variable profiles for the 2D shock tube problem at 
          $t=0.08\cos\alpha$, along the rotated direction $\rm x_1$. 
          The symbols correspond to the CTU-GLM solution whereas the 
          solid lines represent the reference solution. From top to bottom and
          left to right, density, thermal pressure, velocity components and 
          magnetic field components (parallel and perpendicular with respect
          to the $\rm x_1$ direction) are displayed.}
 \label{fig:sod2d}
\end{figure}

Fig \ref{fig:sod2d} shows the primitive variable profiles obtained with the
conservative GLM-MHD scheme. The resulting wave pattern is comprised of two 
outermost fast shocks (at $x_1\sim 0.12$ and $x_1\sim 086$) enclosing 
two slow magneto-sonic waves and a contact mode at $x_1\sim 0.56$.
We see that all discontinuities are captured correctly although 
some spurious oscillations are visible in the transverse velocity profile 
in proximity of the fast shocks. 
Similar features are also evident in the paper by T\'oth \cite{Toth00} 
and with the CT scheme (not shown here).

\begin{figure}[!h]\centering
\includegraphics[width=\textwidth]{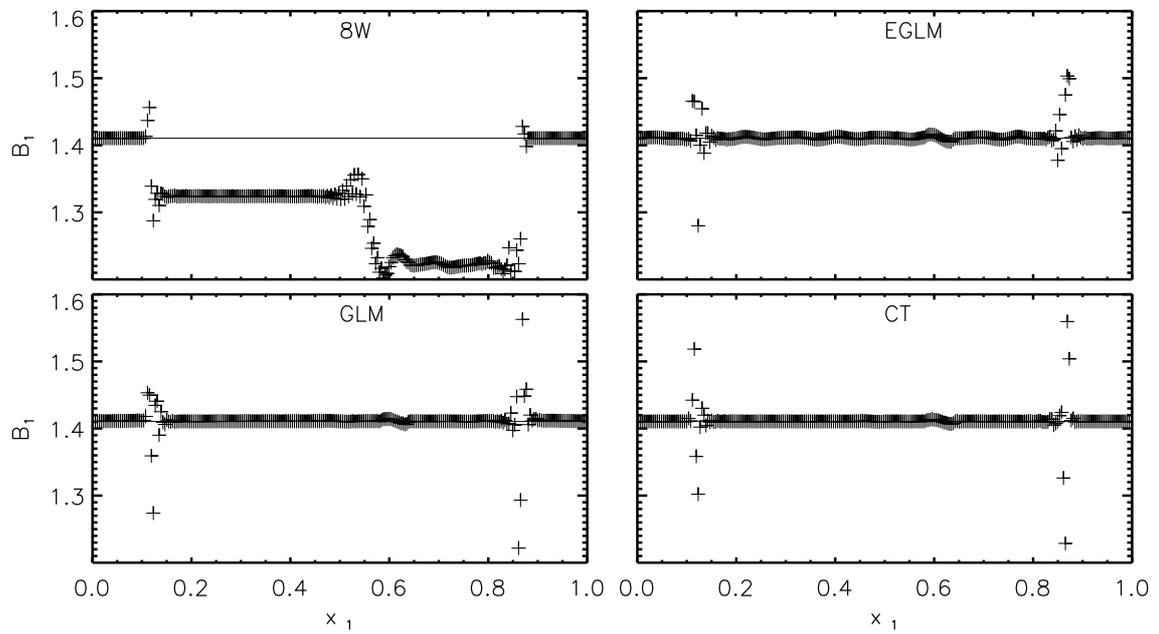}
 \caption{The parallel magnetic field component for the four schemes. 
          Concordantly with the results of \cite{Toth00} the 8 wave formalism 
          fails to capture the correct jumps. 
          This problem is absent in the results of the other schemes and the 
          field component remains close to the expected value $5/\sqrt{4\pi}$
          away from discontinuities.
          Spikes are found in proximity of shock waves and are of the same 
          order of magnitude for GLM, EGLM and CT schemes.}
 \label{fig:sod2d_bn}
\end{figure}

\begin{table}\centering
\caption{One dimensional $L_1 \;(\times 10^{-2})$ norm error for the two-dimensional shock tube.}
\label{table:1}
\centering
\begin{tabular}{c c c c c c c}
\hline\hline
                &  $\rho$  &     $V_1$    &     $V_2$           &  $B_1$   &    $B_2$     &      $p$ \\
\hline  
8W                &$ 2.7 $&$8.6$&$1.9$&${\bf 9.6}$&$6.2$&$94.5$\\ 
CT                &$ 2.6 $&$8.5$&$1.5$&${\bf 0.4} $&$4.7$&$93.0$\\
GLM               &$ 2.6 $&$8.4$&$1.4$&${\bf 0.4} $&$4.3$&$90.5$\\
EGLM              &$ 3.2 $&$8.3$&$1.3$&${\bf 0.4} $&$5.1$&$96.4$\\
\hline
\end{tabular}
\end{table}
We have repeated the same test with the four different schemes described at 
the beginning of this section and compared the results against a 
one-dimensional reference solution obtained at higher resolution 
($1024$ cells) up to $t=0.08$. 
Table \ref{table:1} gives the errors, using the one-dimensional $L_1$ norm, 
of the primitive variables for the 8W, CT, GLM and EGLM schemes. 
While errors in density, velocity and pressure are very similar for all 
schemes, the longitudinal component of the magnetic field ($B_1$) shows 
substantially large deviations with the 8W scheme.
This is further illustrated in Fig \ref{fig:sod2d_bn} where, 
in accordance with \cite{Toth00}, we find that the 8-wave formulation results
in erroneous jump conditions in the normal component of the field.
On the other hand, both the GLM and the non conservative EGLM schemes
behave as well as CT on this particular test without producing 
spurious jump conditions. 

Finally, in the left panel in Fig. \ref{fig:alpha} we plot, 
as a function of $\alpha$, the $L_1$ norm errors in $B_1$ at different 
resolutions, $N_x=128,256,512$, for both the GLM (black) and EGLM 
(red) formulations.
The plots show a weak dependence on the $\alpha$ parameter and 
errors are minimized for $\alpha\approx 0.5$, independently of 
the mesh resolution, for both schemes.
Also, owing to the presence of shock waves, the order of convergence
is approximately one.

\subsubsection{Three-dimensional shock tube}
%
%

For the three dimensional version we follow \cite{GS08}
and set the initial left and right states to
\begin{equation}
\left\{
\begin{array}{lclr}
\vec{V}_L & = & \DS
 \left(1.08,1.2,0.01,0.5,\frac{2}{\sqrt{4\pi}},\frac{3.6}{\sqrt{4\pi}} 
                         \frac{2}{\sqrt{4\pi}},0.95\right)^T 
              & \mathrm{for}\quad  x_1 < 0  \,,\\ \noalign{\medskip}
\vec{V}_R & = & \DS \left(1,0,0,0,\frac{2}{\sqrt{4\pi}},\frac{4}{\sqrt{4\pi}},
                                  \frac{2}{\sqrt{4\pi}},1\right)^T 
              & \mathrm{for}\quad  x_1 > 0  \,,
\end{array}\right.
\end{equation}
where $\vec{V} = \left(\rho,v_1,v_2,v_3, B_1, B_2, B_3, p\right)$ is
the vector of primitive variables.
The coordinate transformation used for the 3D rotation is given by Eq. (\ref{eq:rot_mat})
where the rotation angles $\alpha$ and $\beta$ are chosen in such a way that an 
integer shift of cells satisfies, for any flow quantities $q$, the translational 
invariance expressed by $q({\bf x} + {\bf s})= q({\bf x})$, where 
${\bf s}$ is a Cartesian vector orthogonal to $\vec{x}_1$ and thus 
$x_1({\bf x} + {\bf s}) = x_1({\bf x})$. 
This condition follows from the fact that the solution is a function 
of $x_1$ alone and thus invariant for translations transverse to this direction,
providing a convenient way to assign boundary conditions in the
$(x,y,z)$ system of coordinates.
By choosing $\tan\alpha= -r_1/r_2$ and $\tan\beta = r_1/r_3$ together
with $\vec{s}=(n_x\Delta x,n_y\Delta y, n_z\Delta z)$, 
one can show that the three shift integers $n_x, n_y, n_z$ must obey
\begin{equation}
n_x - n_y \frac{r_1}{r_2} + n_z \frac{r_1}{r_3} = 0 \,,
\end{equation}
where $\Delta x=\Delta y=\Delta z$ has been assumed and
$(r_1,r_2,r_3) = (1,2,4)$ will be used.
The computational domain consists of $[768\times8\times8]$ zones and spans 
$(-0.75,0.75)$ in the $x$ direction while $y,z\,\in [0,0.015625]$.

\begin{table}[!t]\centering
\caption{$L_1\;(\times 10^{-4})$ error for the 3D Shock Tube}
\label{table:2}
\centering
\begin{tabular}{c c c c c c c c c}
\hline\hline

  &  $\rho$  &     $V_1$    &  $V_2$  &   $V_3$     &  $B_1$        &    $B_2$     &  $B_3$ &  $p$ \\

\hline

 8W   &$3.0$&$2.0$&$4.8$&$4.7$&${\bf 3.6}$&$4.5$&$5.1$&$5.0$ \\ 
 CT   &$3.1$&$2.4$&$4.2$&$4.4$& ${\bf 0.5}$&$5.3$&$5.4$&$5.5$ \\
 GLM  &$2.9$&$2.3$&$3.6$&$4.3$& ${\bf 0.5}$&$4.7$&$5.4$&$5.1$ \\
 EGLM &$3.5$&$2.5$&$4.3$&$4.8$& ${\bf 0.5}$&$5.3$&$5.9$&$7.3$ \\
\hline
\end{tabular}
\end{table}

\begin{figure}\centering
\includegraphics[width=\textwidth]{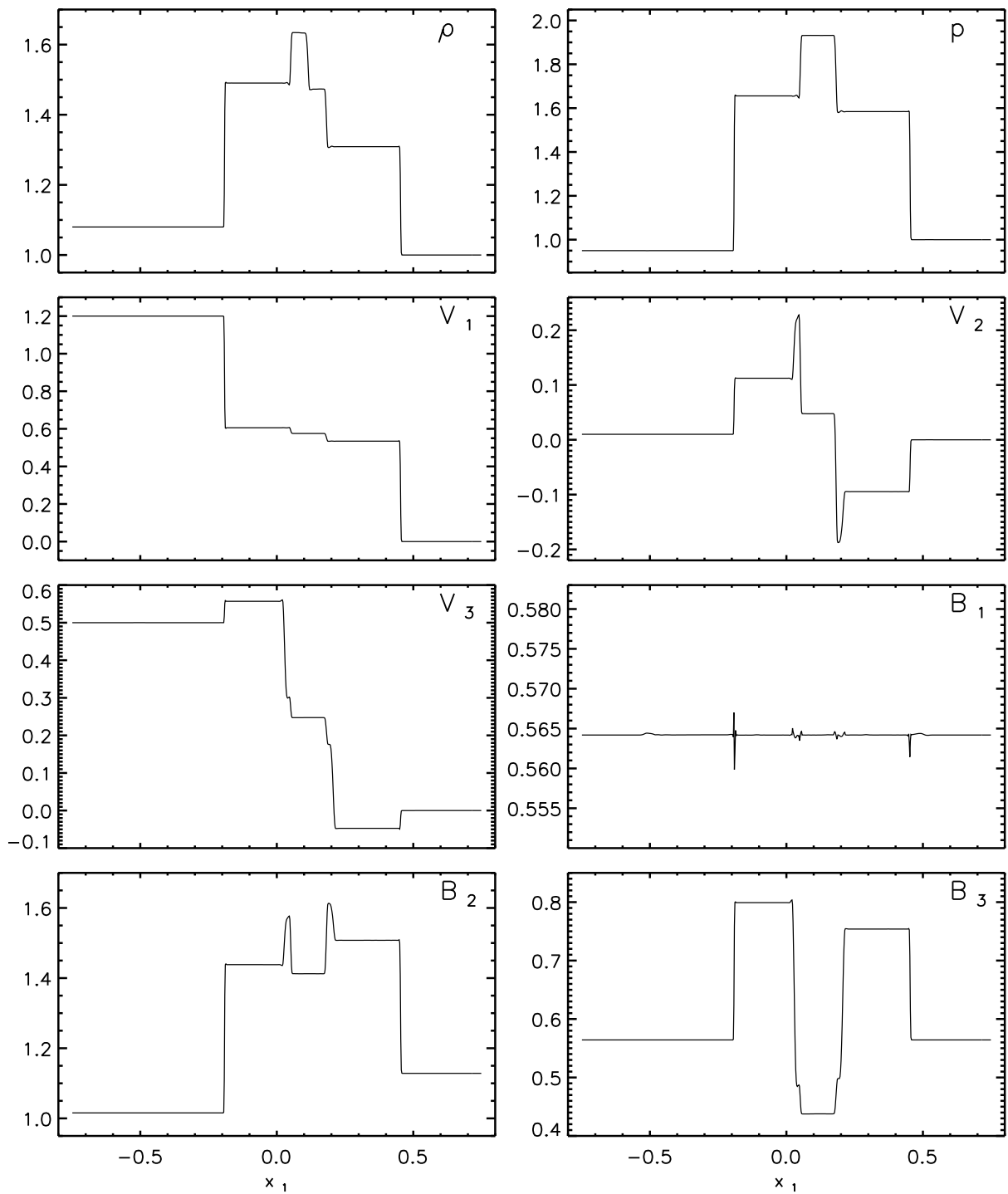}
 \caption{Primitive variable profiles for the 3D shock tube problem at 
          $t=0.02\cos\alpha\cos\gamma$, along the rotated direction $\rm x_1$ .}
 \label{fig:sod_3d}
\end{figure}

In Fig. \ref{fig:sod_3d} we plot the primitive variable profiles for the 
GLM scheme at $t=0.02\cos\alpha\cos\gamma$. 
In accordance with the one dimensional solution (see also \cite{DW94}), we 
observe the formation of a structure involving a contact discontinuity
separating two fast shocks, two slow shocks and a pair of rotational discontinuities.
The three-dimensional integration reproduces the correct behavior of all waves 
and the error in the longitudinal component of the field ($B_1$ in Fig \ref{fig:sod_3d}) 
exhibits small spurious oscillations about the same order of the CT scheme 
(see also, for instance, Fig. 7 in \cite{GS08}).

\begin{figure}\centering
\includegraphics[scale=0.8]{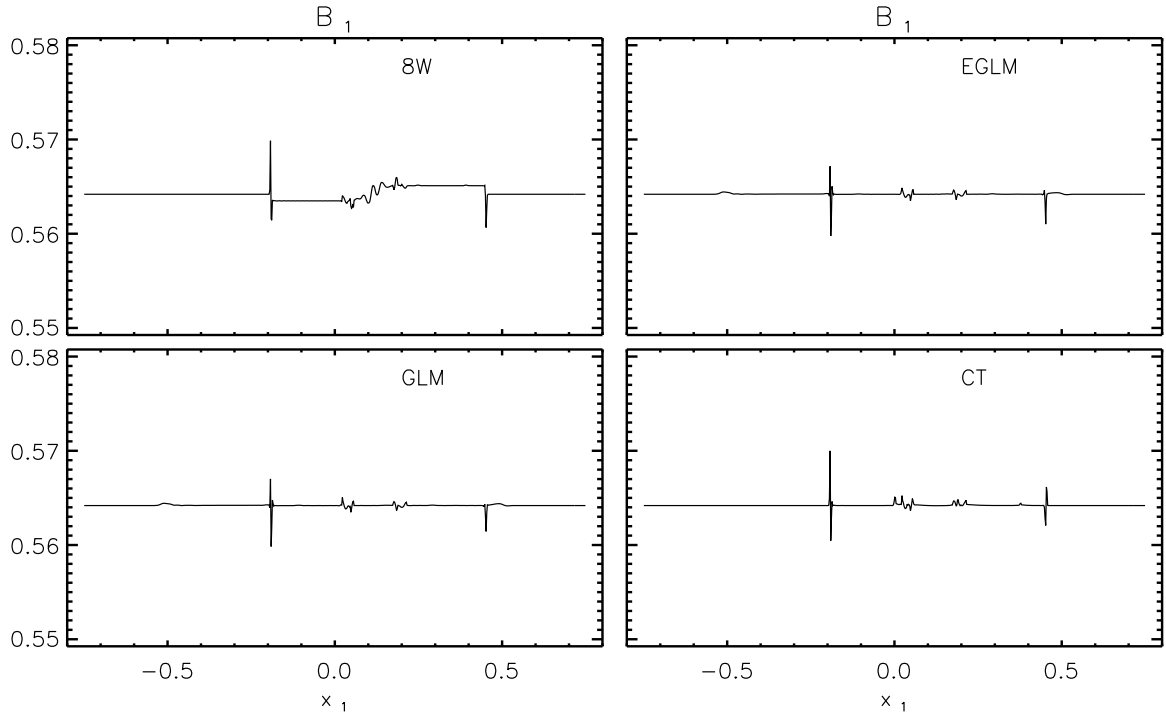}
 \caption{Comparison of the parallel component of the magnetic field
          for the 3D shock tube test.
          As in the 2D case, the error is minimal for all schemes 
          with the exception of the 8-wave formalism. 
          The latter fails to capture correctly the jump but the error
          is less prominent than the 2D case.}
 \label{fig:sod_3d_bn}
\end{figure}

\begin{figure}[!h]
\centering
\includegraphics[width=\textwidth]{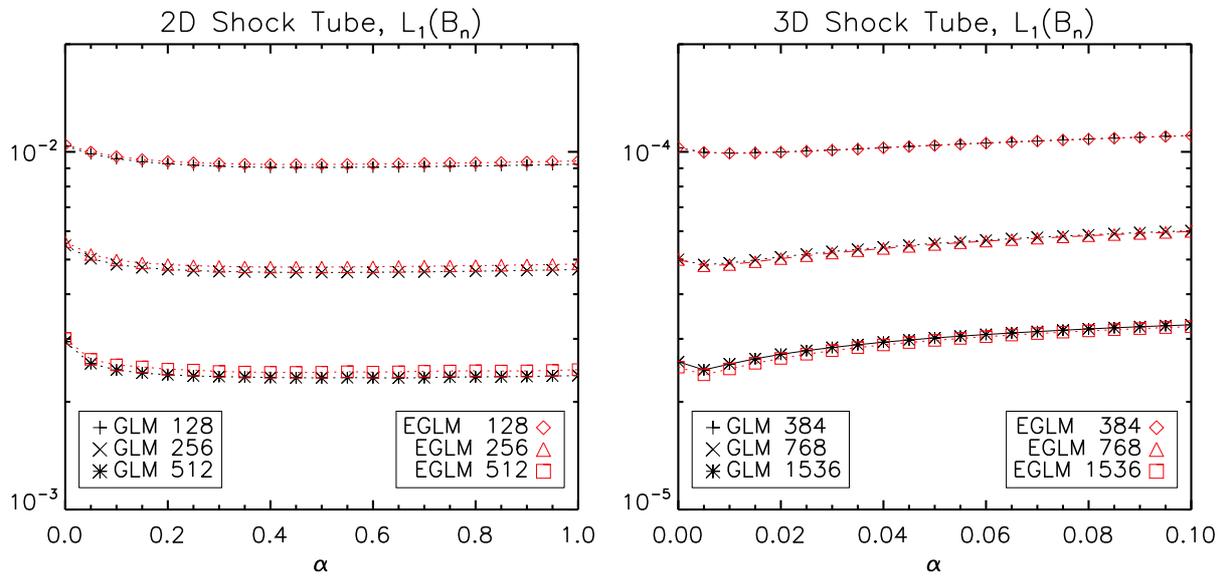}
 \caption{$L_1$ norm errors of $B_1$ (the magnetic field component in the 
          direction orthogonal to the initial discontinuity) as functions of 
          $\alpha = \Delta hc_h/c_p^2$ for the 2D (left) and 3D (right) shock 
          tube problem. The different symbols correspond to computations 
          carried at different mesh resolutions: $N_x=128,256,512$ (in 2D) 
          and $N_x=384,768,1536$ in 3D. 
          Black and red symbols refer to results obtained with the GLM and
          EGLM formulations, respectively. The Courant numbers were
          $0.8$ and $0.4$ for 2 and 3D computations, respectively.}
 \label{fig:alpha}
\end{figure}

A quantitative estimate of the error (using the one-dimensional $L_1$ norm 
error) is obtained by comparing the three-dimensional results with a 
one-dimensional reference solution computed on $1024$ zones until $t=0.02$. 
The comparison, extended to the four selected integration schemes, 
is given in Table \ref{table:2}. 
We notice that the CT, GLM and EGLM schemes all 
yield errors of the same order of magnitude (typically $10^{-4}$).
Beware that these computations may be susceptible to small variations depending on 
implementation details (e.g. limiter, Courant number, etc.) and thus give a 
representative estimate of the error.
For instance, the implementation of the CTU-CT scheme in the PLUTO code 
\cite{Mignone07} is similar, although not exactly equivalent, to that of 
\cite{GS08} who instead use piecewise parabolic reconstruction. 
Nevertheless, we have ascertained that the 8W scheme always performs the worst 
and the discrepancy becomes particular evident by looking at the longitudinal 
component of the field where the 8W scheme yields, once again, incorrect 
(although smaller than the previous 2D case) jumps.
This is better illustrated in Fig. \ref{fig:sod_3d_bn}, where we compare the 
profiles of $B_1$ for the four selected numerical schemes.
We stress that, despite its non-conservative character, the EGLM formulation 
does not seem to produce incorrect jump conditions or wrong shock propagation
speeds.

A resolution study, shown in the right panel of Fig \ref{fig:alpha}, demonstrates 
that errors produced by the GLM and EGLM formulations are very much 
comparable and only weakly dependent on the $\alpha$ parameter.
Both schemes report a minimum at $\alpha\approx 0.005-0.01$
regardless of the resolution, and the inferred order of convergence 
is approximately one as expected for solutions involving 
shock waves.

\subsection{Magnetic Field Loop Advection}
%
%
%
%

This problem consists of a weak magnetic field loop being 
advected in a uniform velocity field. 
Since the total pressure is dominated by the thermal contribution, 
the magnetic field is essentially transported as passive scalar.

\subsubsection{Two-dimensional advection}
%
%

Following \cite{GS05, FHT06,LD09}, we employ a periodic computational box defined 
by $x\in[-1,1]$ and $y\in[-0.5,0.5]$ discretized on $N_x\times N_x/2$ grid cells 
($N_x=128$). Density and pressure are initially constant and equal to $1$. 
The velocity of the flow is given by $\vec{v} = (V_0\cos\alpha, V_0\sin\alpha, 1)$ 
with $V_0 = \sqrt{5}$, $\sin \alpha = 1/\sqrt{5}$ and $\cos \alpha = 2/\sqrt{5}$. 
The magnetic field is defined through its magnetic vector potential as 
\begin{equation}  
A_z = \left\{ \begin{array}{ll}
    A_0(R-r) & \textrm{if} \quad r \leq R \,, \\ \noalign{\medskip}
    0   & \textrm{if} \quad r > R \,,
  \end{array} \right.
\end{equation}     
where $A_0 = 10^{-3}$, $R=0.3$ and $r=\sqrt{x^2 + y^2}$. 
The simulations are allowed to evolve until $t=2$ ensuring the
crossing of the loop twice through the periodic boundaries. 

\begin{figure}\centering
\includegraphics[width=0.9\textwidth]{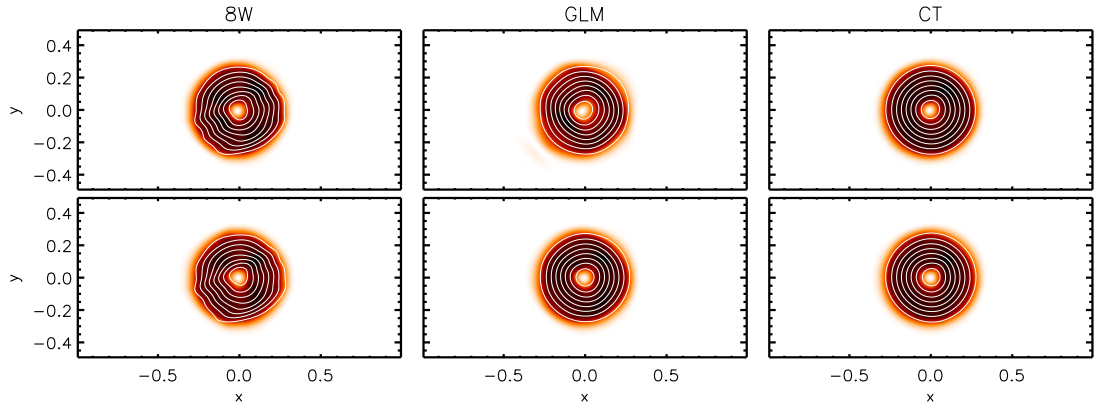}
 \caption{From left to right: magnetic energy density for the 2D field loop 
  problem at $t=2$ for the 8W, GLM and CT schemes.
  Results have been computed with CFL numbers of $0.8$ (top) and 
  $0.4$ (bottom). Overplotted are 9 isocontours of $A_z$, between $10^{-5}$ 
  and $10^{-3}$.}
 \label{fig:fl2d}
\end{figure}

In Fig. \ref{fig:fl2d} we show the magnetic energy density for the 8W, GLM and CT 
schemes using $C_a=0.8$ (top) and $C_a=0.4$ (bottom), along with the field lines 
shape. The circular shape of the loop is best preserved with the CT and GLM 
schemes while some distortions are visible using the 8 wave formulation. Using 
$C_a = 0.4$ with the GLM scheme yields slightly better results, while the CT does 
not seem to be affected by the choice of the Courant number.

\begin{figure}\centering
\includegraphics[width=0.9\textwidth]{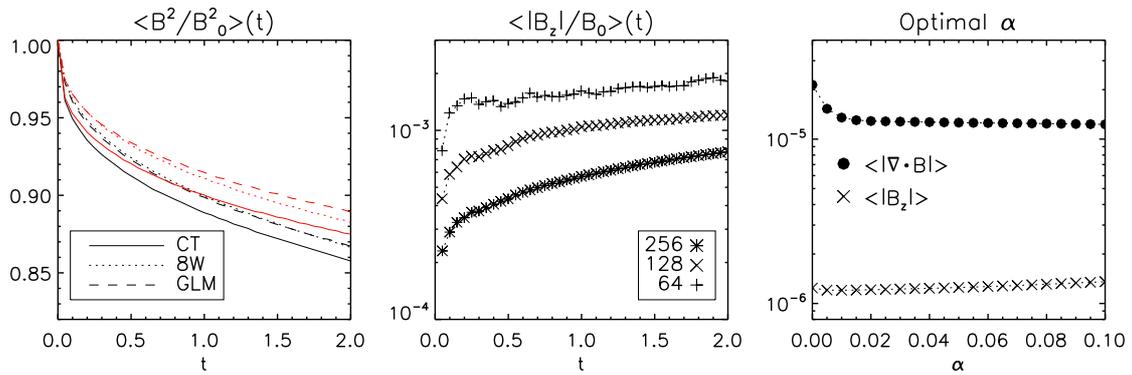}
 \caption{Leftmost panel: time evolution of the volume-integrated 
  magnetic energy density (normalized to its initial value) for the 2D field 
  loop advection problem. 
  The black and red lines correspond, respectively, to computations carried 
  with $C_a=0.4$ and $C_a=0.8$.
  Middle panel: volume-averaged value of $|B_z|$ (normalized to 
  the initial value $B_0=10^{-3}$) as a function of time for three different 
  grid resolutions ($256, 128$ and $64$ corresponding to stars, "x" and plus 
  signs). 
  Rightmost panel: volume-averaged values of $|\nabla\cdot\vec{B}|$
  and $|B_z|$ for different values of the $\alpha$ parameter controlling 
  monopole damping at the resolution $N_x=128$ points.}
 \label{fig:fl2d_energy}
\end{figure}

The time-history of the magnetic energy density (left panel in Fig 
\ref{fig:fl2d_energy}) reveals that the numerical dissipation is essentially 
similar for all schemes, being smaller at larger Courant numbers. 
At the quantitative level, our results are similar and in good agreement with 
those of other investigators (e.g., .\cite{GS05,FHT06,LD09}).

The ability of the GLM scheme in preserving the divergence-free condition is 
monitored by checking the growth of $B_z$ in time: owing to a non-vanishing $z$ 
component of velocity, in fact, we expect $B_z$ to grow in time with a rate
$\propto v_z\nabla\cdot\vec{B}$ as seen from the induction equation.
In the middle panel of Fig. \ref{fig:fl2d_energy} we plot the volume-averaged 
value of $B_z$ as a function of time for $N_x=64,128,256$.
The nominal value is $\sim 10^{-3}$ of the initial field strength, decreasing 
with resolution.
Notice that the observed order of convergence is $\sim 0.6-0.7$ 
and thus sub-linear as expected for a linearly degenerate wave
in Godunov-type schemes, in accordance with the results of \cite{BAR08}.

Computations carried with different values of $\alpha$ reveal that divergence 
errors are minimized for $\alpha \gtrsim 0.01$ while errors in $B_z$ become 
smallest for $\alpha \approx 0.01$ (right panel in Fig \ref{fig:fl2d_energy}).
Despite this may generate some ambiguities in prescribing an optimal
$\alpha$ value, however, we see that its choice does not significantly 
affect the error and thus constitutes a minor effect on the solution.

\subsection{Three-dimensional field loop advection}
%
%


A three-dimensional extension can be obtained by rotating the previous 2D 
configuration around one axis using the coordinate transformation given 
by Eq. (\ref{eq:rot_mat}) with $\alpha=0$ and $\gamma = \tan^{-1}1/2$,
see \cite{GS08}.
Even though the loop is rotated only around one axis, the velocity profile 
$(v_x,v_y,v_z)=(1,1,2)$ makes the test intrinsically three-dimensional.
We consider the computational box $-0.5\le x \le 0.5$, $-0.5\le y \le 0.5$,
$ -1.0\le z \le 1.0$, resolved on a $N\times N\times 2N$ grid. 
Boundary conditions are periodic in all directions.

\begin{figure}\centering
\includegraphics[width=\textwidth]{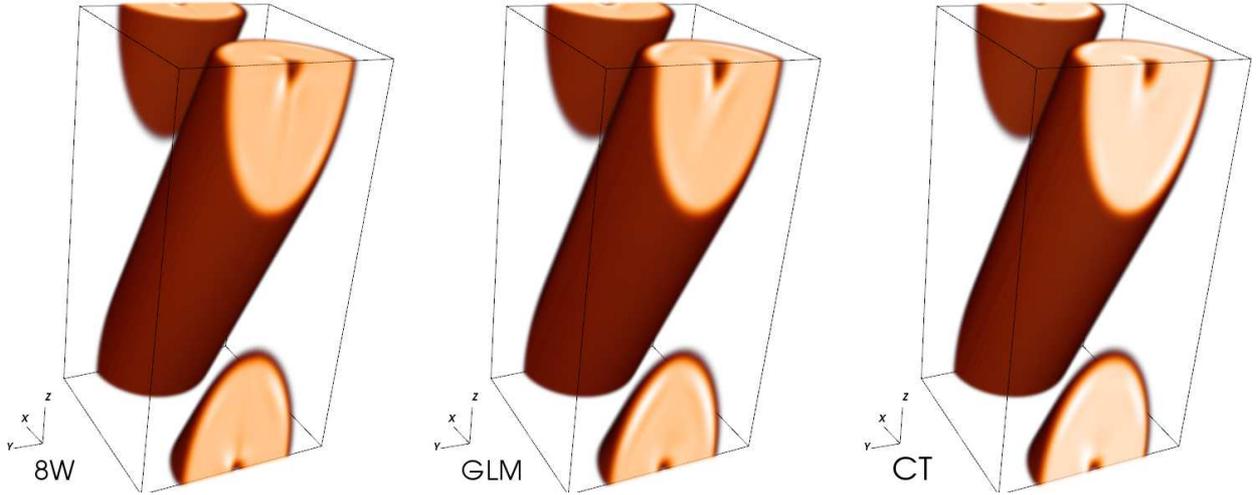}
 \caption{Magnetic energy density for the 3D field loop problem
          at $t=1$ at the resolution of $128\times 128\times 256$. 
          From left to right: results obtained with the 8W, GLM and 
          CT schemes.}
 \label{fig:f7}
\end{figure}

\begin{figure}\centering
\includegraphics[width=0.9\textwidth]{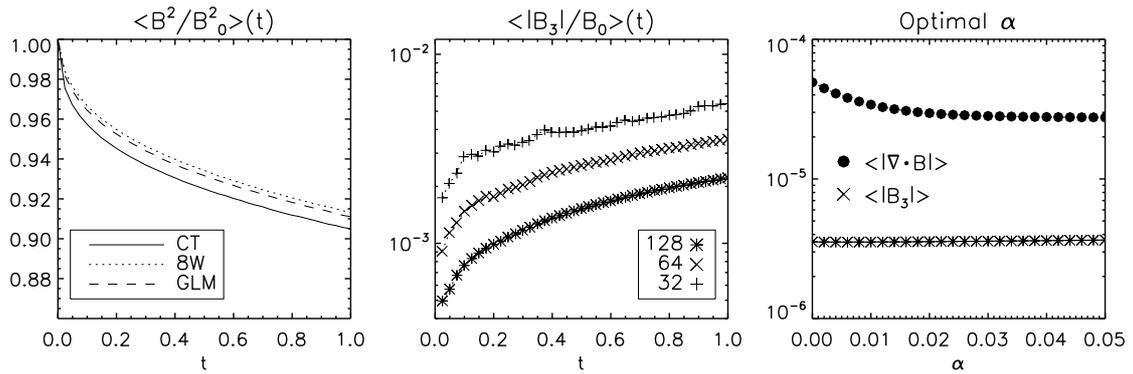}
 \caption{Same as Fig \ref{fig:fl2d_energy} for the 3D field loop
  advection test.
  From left to right: time history of the (normalized) volume-integrated
  magnetic field energy, (normalized) average value of $|B_3|$ (magnetic field 
  component orthogonal to the original 2D plane) and volume averages of 
  $|\nabla\cdot\vec{B}_3|$ and $|B_3|$ as functions of the $\alpha$
  parameter.}
 \label{fig:f8}
\end{figure}

A three-dimensional rendering of the magnetic energy density is shown in Fig. 
\ref{fig:f7} for the selected schemes while relevant quantities are plotted in the
three panels of Fig \ref{fig:f8}. 
All schemes show a similar amount of numerical dissipation, in agreement with 
the results of \cite{GS08}.

As for the 2D case, it is useful to check the growth of the magnetic field
component $B_3=(-B_x+2B_z)/\sqrt{5}$ orthogonal to the original $(x_1,x_2)$ plane
where the loop is two-dimensional.
Analytically, the magnetic field component in this direction is a 
trivial constant of motion since
\begin{equation} 
 \pd{B_3}{t} = v_{3}\left(\pd{B_1}{x_1} + \pd{B_2}{x_2}\right) = 0\,.
\end{equation}
The numerical integration in the rotated $(x,y,z)$ Cartesian frame, 
however, preserves this condition only to some accuracy which strongly 
reflects the ability of the scheme in controlling the divergence-free 
constraint (this is true for all presented numerical methods).
The middle panel in Fig \ref{fig:f8} shows the volume-integrated value 
of $|B_3|$, normalized to the initial field
strength $B_0 = 10^{-3}$ for three different resolutions $N=32,64,128$.
Our results reveal that the value of $B_3$ grows slowly in time while remaining
reasonably small.
The convergence rate ($\approx 0.6-0.7$) is approximately the same as 
the one observed in the 2D case.
 
The dependency on $\alpha$ is illustrated in the right panel Fig \ref{fig:f8} 
showing that  divergence errors are progressively reduced for $\alpha \gtrsim 0.03$ 
although this has very little effect on the growth of $B_3$.

\subsection{Two-dimensional Rotor problem}
%
%

The rotor problem consists of a dense disk rotating in a static medium 
threaded by an initially uniform magnetic field. 
As the rotor spins, the magnetic field gets wrapped around the disk
creating torsional Alfv\'en waves, stemming from the rotating disk
and moving towards the surrounding gas. 
This interaction slows down the disk by extracting angular momentum. 
On the other hand, the build-up of magnetic pressure around the rotor 
causes its compression.

We initialized the problem on the Cartesian box $x,y\in[-\HALF,\HALF]$ 
with outflow boundary conditions and use $400^2$ grid points.
The primitive variable profiles at the beginning of the simulation are given by 
\begin{equation}  
\left(\rho, v_x, v_y\right) = \left\{ \begin{array}{ll}
\DS \left(10, -\omega y, \omega x\right) & \textrm{if $r \leq r_0$} \,, \\ \noalign{\medskip}
\DS \left(1 + 9 f, -f\omega y \frac{r_0}{r}, f\omega x \frac{r_0}{r}\right) & \textrm{if $r_0<r<r_1$} \,,\\ \noalign{\medskip}
\DS \left(1, 0, 0\right)         & \textrm{if $r \geq r_1$}  \,,   
\end{array} \right. 
\end{equation}  
where $\omega=20$, $r_0 = 0.1$, $r_1 = 0.115$, $r=\sqrt{x^2 + y^2}$ and 
the taper function is $f=(r_1 - r)/(r_1 - r_0)$. Thermal pressure 
is initially uniform and equal to one ($\Gamma = 1.4$ is used). 
The magnetic field has only one non-vanishing component, $B_x=5/\sqrt{4\pi}$.

\begin{figure}\centering
\includegraphics[scale=0.8]{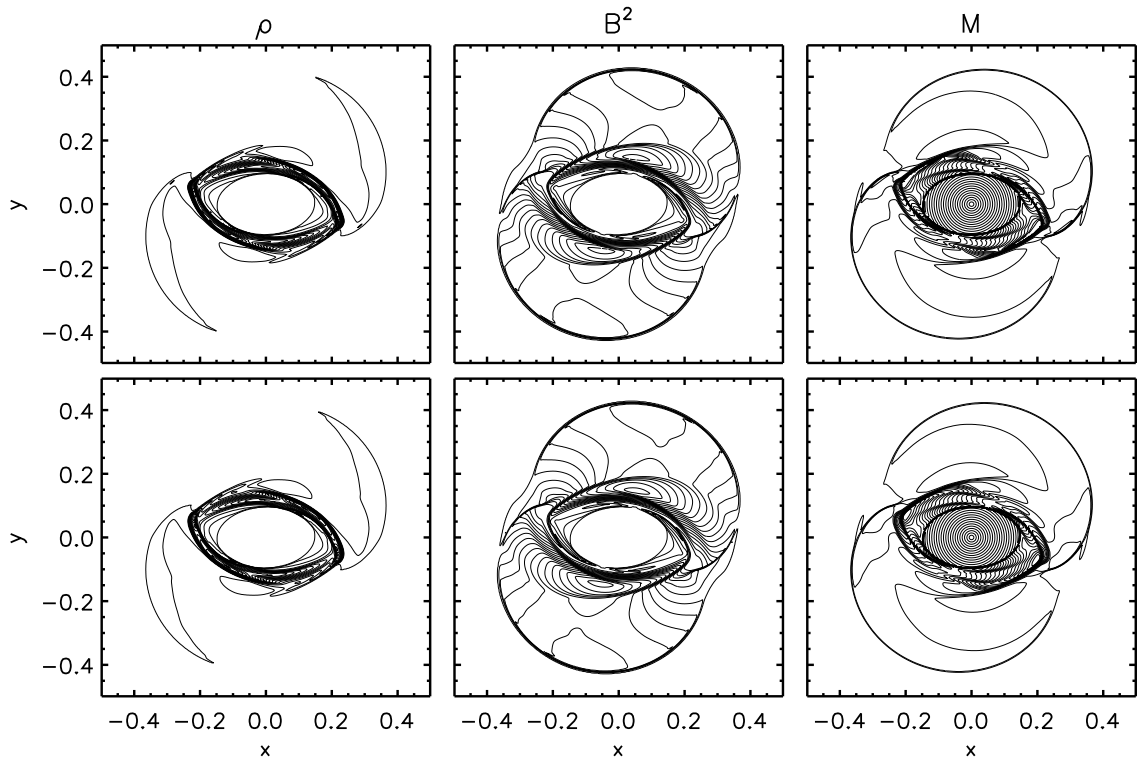}
 \caption{Density, magnetic energy and sonic Mach number for the 
          rotor problem at $t=0.15$ obtained with the GLM (upper panels) 
          and the CT (lower panels) methods. $20$ levels are displayed,
          the range of which is $0.5\le\rho\le13$, $0.04\le B^2\le5.2$ and
          $0\le M \le 4$}
 \label{R1}
\end{figure}

\begin{figure}\centering
\includegraphics[scale=0.8]{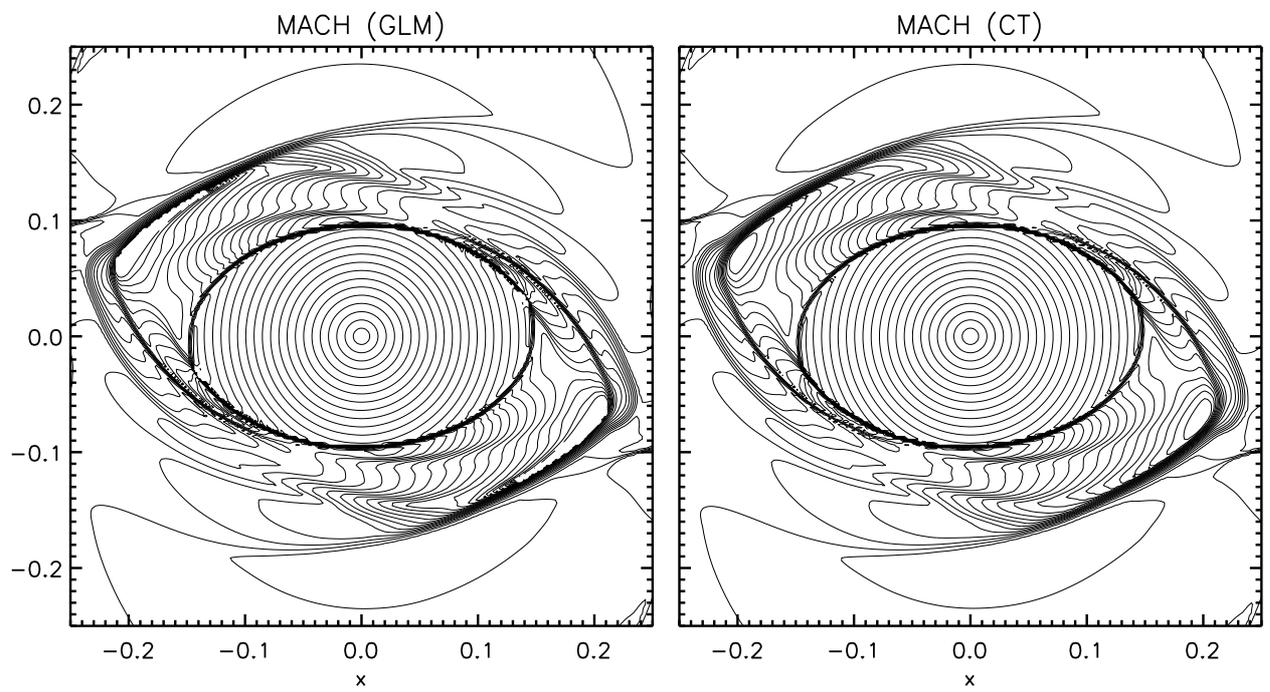}
 \caption{A zoom in the central region of the rotor problem at time $t=0.15$,
          showing $20$ levels ($0\le M \le 4$) of contour profiles of the sonic Mach number. 
          Results for the GLM and CT schemes are shown on the left 
          and right panels, respectively.}
 \label{R2}
\end{figure}

The maps of density, magnetic energy and sonic Mach number are displayed in 
Fig. \ref{R1} at $t=0.15$ for the GLM and the CT schemes, when
the torsional Alf\'en waves have almost reached the outer boundaries.
The strength of the scheme is also measured by its ability to preserve the 
circular shape of the sonic Mach number profile in the central region, an essential 
feature of the solution, \cite{LD09}.
This is better shown in Fig. \ref{R2} where an enlargement of the central 
region reveals that the GLM and CT schemes have developed extremely similar 
Mach number contours and the absence of spurious peaks (that would be 
caused by pressure undershoots) advocates towards the validity of the 
scheme.
 
\subsection{Three Dimensional Blast Wave}
%
%

\begin{table}[!t]\centering
\caption{Parameter sets used for the first and second versions of the three-dimensional 
         blast wave problem.}
\label{tab:blast_ic}
\centering
\begin{tabular}{ccccccc}
\hline\hline
  & $p_{\rm in}$ & $p_{\rm out}$ & $B_0$ & $\theta$  & $r_0$   & $t_{\rm stop}$  \\
\hline\noalign{\medskip}
 Test 1 & $10^2$ &   $1$         & $10$      & $\pi/4$ & $0.125$ & 0.02 \\ \noalign{\medskip}
 Test 2 & $10^4$ &   $1$         & $100$     &   $0$   & $0.1$   & $2.5\cdot 10^{-3}$ \\
\hline
\end{tabular}
\end{table}
The MHD blast wave problem has been specifically designed to show the scheme 
ability to handle strong shock waves propagating in highly magnetized 
environments, see for instance \cite{ZMC94,  BS99, Ziegler04, GS08, LD09}.
Depending on the strength of the magnetic field, it can become a rather arduous 
test leading to unphysical densities or pressures if the divergence-free 
condition is not properly controlled and the scheme does not introduce adequate 
dissipation across oblique discontinuous features.
Here, we consider a three-dimensional configuration on the unit cube 
$[-1/2,1/2]^3$ discretized on $200^3$ computational zones.
The medium is initially at rest ($\vec{v}=\vec{0}$) and threaded by a constant 
uniform magnetic field lying in the $xz$ plane and forming an angle $\theta$ with 
the vertical $z$ direction, 
$\vec{B}=B_0\left(\sin\theta\hat{x}+\cos\theta\hat{z}\right)$.
A spherical region of high thermal pressure is initialized, 
\begin{equation}
 p = \left\{\begin{array}{ll}
  p_{\rm in}  & \quad\textrm{for}\quad \sqrt{x^2+y^2+z^2} < r_0 \,,\\ \noalign{\medskip}
  p_{\rm out} & \quad\textrm{otherwise} \,.
\end{array}\right.
\end{equation}
We consider two different versions of the same test problem with parameters 
given in Table \ref{tab:blast_ic}.
In the first one, taken from \cite{GS08}, the field forms an angle $\theta=\pi/4$ 
with the $z$ axis and the largest magnetization achieved outside
the sphere is $\beta = 2p_{\rm out}/\vec{B}^2 = 2\cdot 10^{-2}$.
In the second version, we follow \cite{Ziegler04} and adopt a a larger field 
strength (with $\theta = 0$) yielding a more severe configuration with 
$\beta = 2\cdot 10^{-4}$.

\begin{figure}\centering
\includegraphics[width=0.95\textwidth]{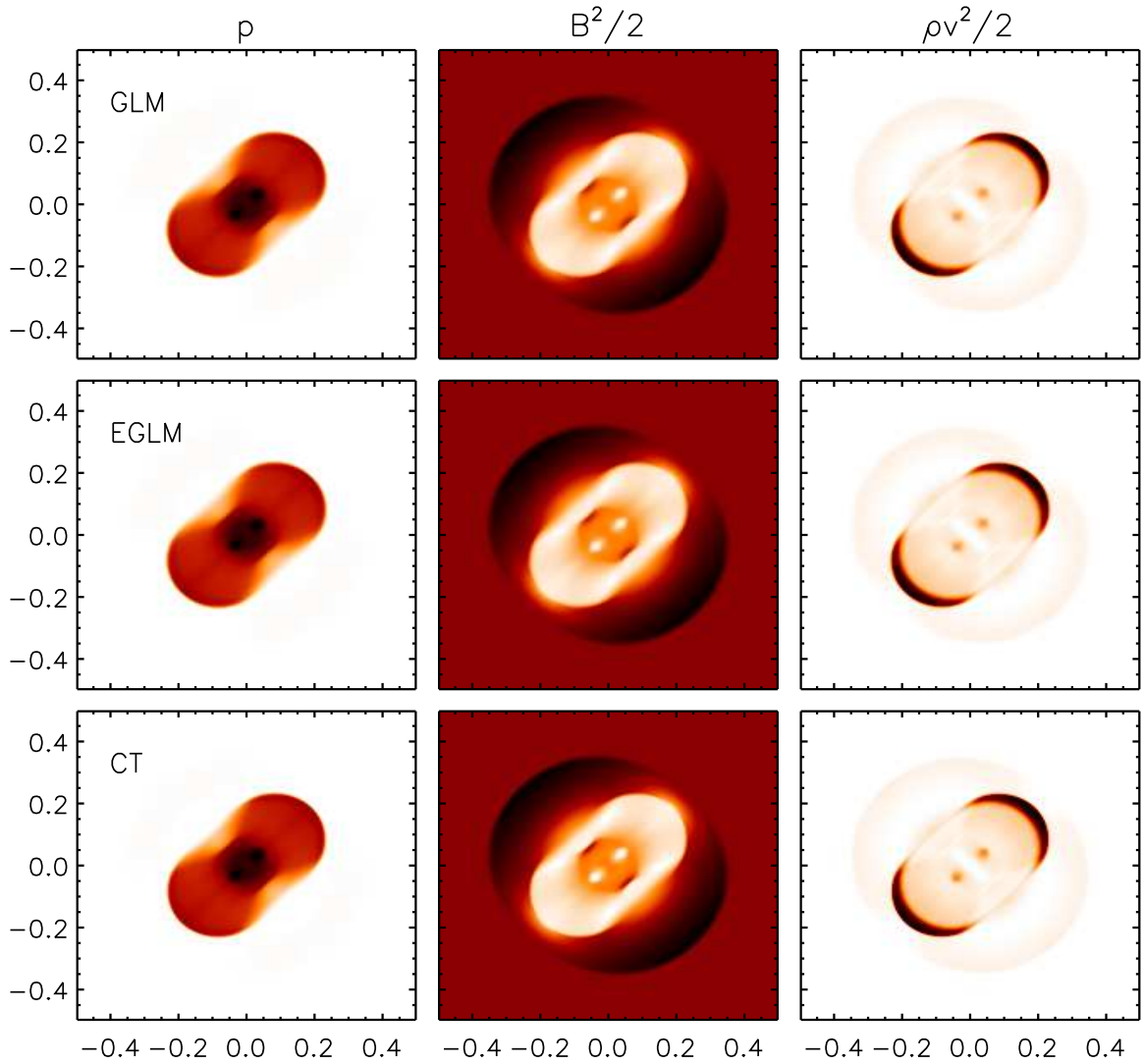}
 \caption{Two dimensional cuts in the $xz$ plane of gas pressure, magnetic and 
          kinetic energy densities for the GLM (top), EGLM (middle) and CT 
          (bottom) schemes, at $t=0.02$ for the first blast wave problem. 
          Pressure values range from $1.0$ (white) to $42.4$ (black). 
          The magnetic energy ranges from $25.2$ (white) to $64.9$ (black) 
          while the kinetic energy density spans from $0.0$ (white) to 
          $33.1$ (black). }
 \label{fig:blast1a}
\end{figure}

The over-pressurized spherical region sets a blast wave delimited by an outer 
fast forward shock propagating (nearly) radially, see Fig \ref{fig:blast1a} and 
\ref{fig:blast2}.
Magnetic field lines pile up behind the shock in the direction transverse to the 
initial field orientation ($\theta = \pi/4$ and $\theta=0$ for the two cases)
thus building a region of higher magnetic pressure.
In these regions the shock becomes magnetically dominated and only weakly 
compressive ($\delta\rho/\rho\sim 1.2$ in both cases).
The inner structure is delimited by an oval-shaped slow shock adjacent to a 
contact discontinuity and the  two fronts tend to blend together as the 
propagation becomes perpendicular to the field lines. 
The magnetic energy increases behind the fast shock and 
decreases downstream of the slow shock.
The resulting explosion becomes highly anisotropic and magnetically confined.

\begin{figure}\centering
\includegraphics[width=0.95\textwidth]{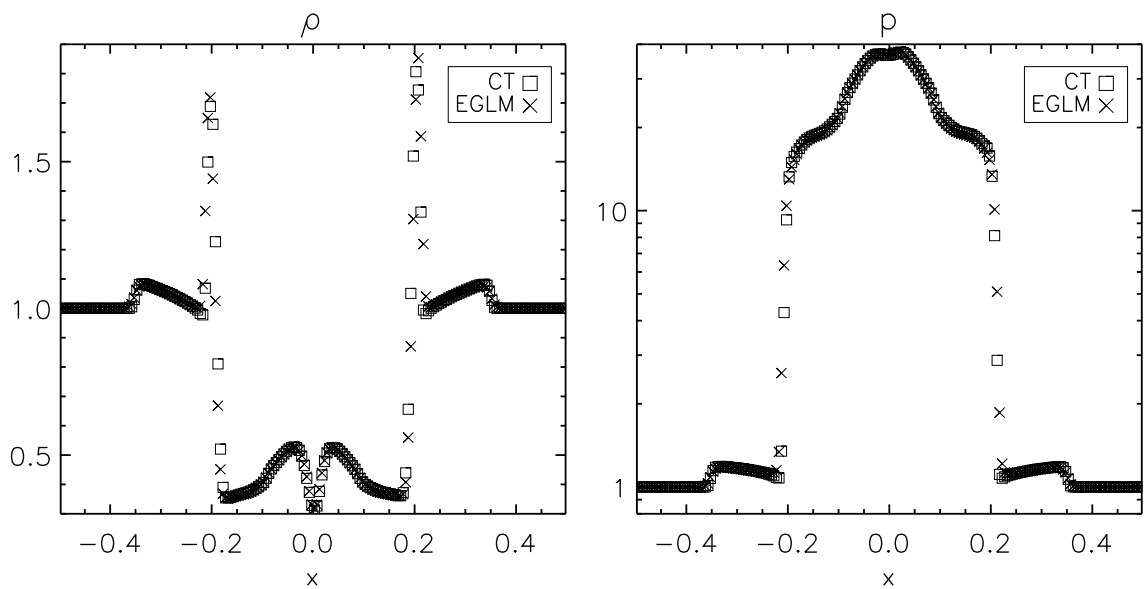}
 \caption{Density and pressure profiles, the latter on logarithmic scale, 
          along $\rm{x}$ at $\rm{y,z} = 0$, at time $t=0.02$.
          Results obtained with the CT and EGLM schemes are shown using box 
          and cross symbols, respectively.}
 \label{fig:blast1b}
\end{figure}

\begin{figure}\centering
\includegraphics[width=0.95\textwidth]{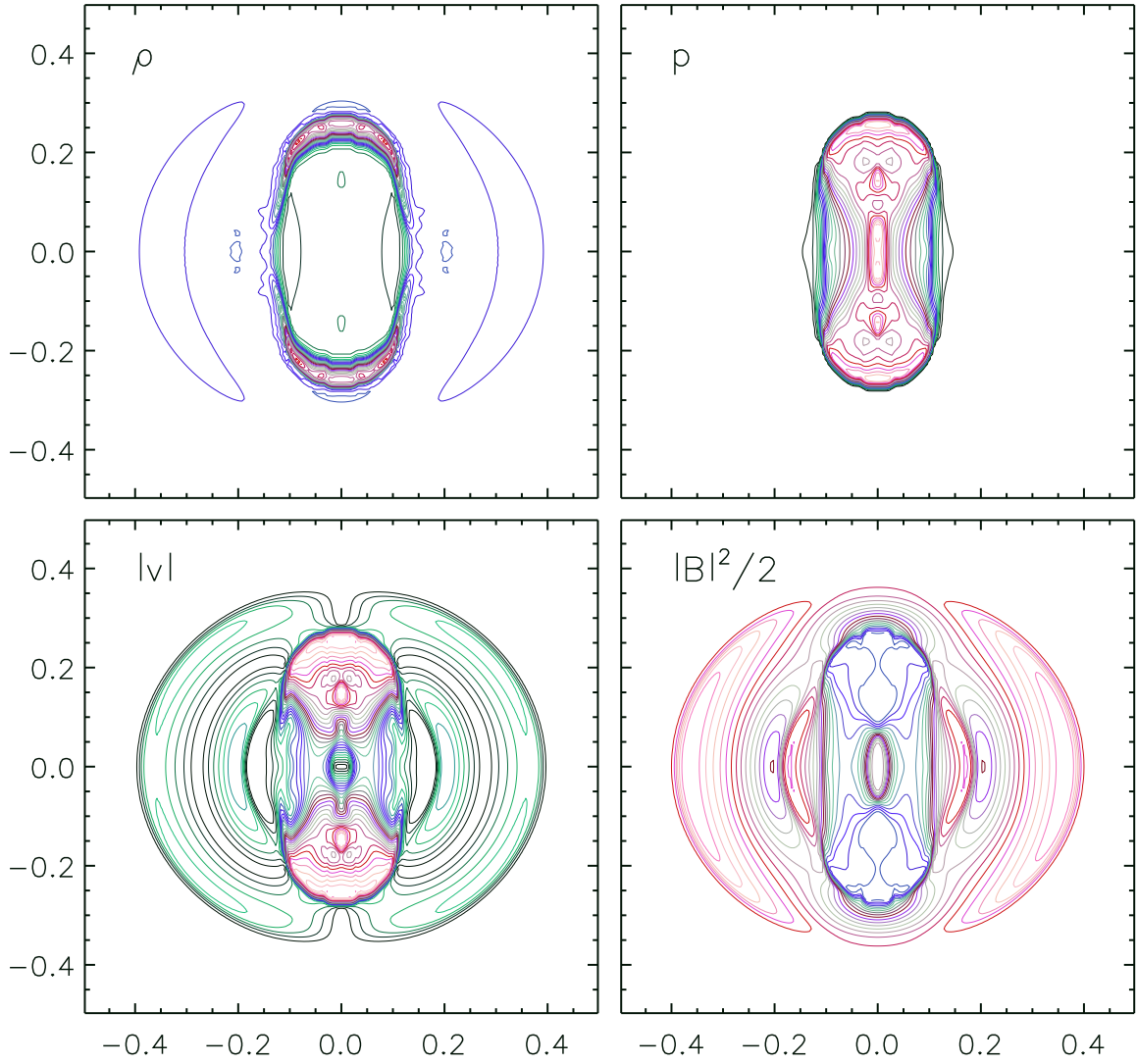}
 \caption{Density, pressure, velocity and magnetic energy contours (30 levels) for the 
          CTU-EGLM scheme at $t=2.5\cdot 10^{-3}$ in the $xz$ plane.
          Density values range from $0.18$ to $3.2$ while pressure spans from $0.9$ to $2290$. 
          The absolute value of velocity ranges from $0.0$ to $47$ while the magnetic energy spans 
          from $2817$ to $5932$.}
 \label{fig:blast2}
\end{figure}

Computed results for the first configuration are shown in Fig \ref{fig:blast1a},
where we display linearly scaled maps of gas pressure, magnetic and kinetic energy
densities for the GLM scheme (top), EGLM (middle) and CT schemes (bottom).
The computations are in excellent agreement and no noticeable difference can be 
discerned from the images. 
Moreover, our results favorably compare to those of \cite{GS08}. 
To further ascertain the validity of the non-conservative EGLM scheme, 
we plot, in Fig \ref{fig:blast1b}, one-dimensional slices (along the $x$ direction 
in the $yz$ mid-plane) showing the density and pressure obtained 
with the EGLM and CT integrations.
 
Computations for the second configuration could be obtained only with the EGLM 
scheme, since the CT scheme failed even with a minmod limiter ($\beta = 1$ in 
Eq. \ref{eq:limiter}). 
In Fig. \ref{fig:blast2} we plot contour levels for density, pressure, velocity 
and magnetic energy. These results comply with those of \cite{Ziegler04} who used 
a CT scheme together with a Runge-Kutta time stepping and an HLL Riemann solver.
They also share similarities with the 2D strong field case discussed in 
\cite{LD09} who used a different implementation of the CT scheme.
Partially owing also to the increased resolution ($200^3$ instead of $144^3$) our 
CTU-EGLM algorithm shows considerably reduced numerical diffusion while being 
robust in keeping sharp profiles of the discontinuities. 

\section{Conclusions}
\label{sec:conclusions}
%
%
%

A second-order, cell-centered numerical scheme for the solution of 
the MHD equations in two and three dimensions has been proposed.
Fully unsplit integration resorts to the Corner Transport Method of Colella 
\cite{Colella90} and the divergence-free condition is controlled by using a 
constrained formulation of the MHD equations where the induction equation 
is coupled to a generalized Lagrange multiplier (GLM, \cite{Dedner02}). 
The system is hyperbolic, easy to implement and does not require expensive 
cleaning projection steps associated with the solution of elliptic problems.
The GLM scheme is fully conservative in mass, momentum, energy and magnetic
induction, although we have also considered a slightly modified variant 
(EGLM) which infringes momentum and energy conservation.

In order to assess the reliability and accuracy of the schemes we have performed 
a number of code benchmarks on standard two- and three-dimensional MHD test 
problems.
Results have been compared with two different numerical schemes: a 
non-conservative cell-centered method based on the 8-wave formulation
(8W, \cite{Powell99}) and the constrained transport (CT) method where the 
magnetic field has a staggered collocation. 
Both the GLM and EGLM schemes give excellent results in terms of accuracy and 
robustness and do not show, in the tests presented here, any evidence for incorrect 
jump conditions or wrong wave propagation, as found for the eight wave formulation
(in agreement with T\'oth \cite{Toth00}). 
This has been verified on problems involving discontinuous waves and holds true 
for both the conservative GLM formulation \emph{and} the EGLM variant which 
breaks momentum and energy conservation.
In this perspective, our results seem to indicate that the presence of source 
terms in the equations does not necessarily lead to erroneous jumps.
Instead, we have found the non-conservative formulation to be more robust for 
problems involving the propagation of oblique strongly magnetized shocks.
Although, this behavior may be attributed to discretization, 
such a study is beyond the scope of the present paper.
The comparison has also revealed an excellent quantitative agreement with the 
CTU-CT scheme (in the version of \cite{GS05,GS08}) showing errors with comparable 
magnitude and similar order of convergence while retaining the desired robustness 
and stability.

For these reasons, we believe that the proposed CTU-GLM and CTU-EGLM schemes
provide excellent competitive alternatives to modern staggered-mesh algorithms 
while being considerably easier and more flexible in their implementations.
Owing to the cell-centered collocation of all of the flow fields, 
the CTU-GLM scheme can be easily generalized to resistive MHD, adaptive
and/or unstructured grids and to higher than second-order spatially-accurate 
numerical schemes. Some of these issues will be presented in forthcoming papers.




\appendix
%
%
\section{Characteristic Decomposition of the GLM-MHD Equations}
\label{app:eigenv}
%
%
%
%

The $9\times 9$ matrix $\tens{A}_x$ of the primitive MHD equations
introduced in \S\ref{sec:normal} can be decomposed as 
$\tens{A}_x = \tens{R}\tens{\Lambda}\tens{L}$ where 
$\tens{\Lambda} = {\rm diag}(\lambda^k)$  contains the eigenvalues 
(see Eq. \ref{eq:eigenvalues}) while the rows of $\tens{L}$ and columns of 
$\tens{R}$ are the corresponding left and 
right eigenvectors of $\tens{A}_x$, respectively.
Adopting the scaling of \cite{Powell99} we define 
\begin{equation}
\alpha_{f}^2 = \frac{\alpha^2-c_{s}^2}{c_{f}^2-c_{s}^2}, \quad\quad \alpha_{s}^2 = \frac{c_{f}^2 - \alpha^2}{c_{f}^2-c_{s}^2}
\end{equation}
and
\begin{equation}
\beta_y = \frac{B_y}{\sqrt{B_{y}^2+B_{z}^2}}, \quad\quad \beta_z = \frac{B_z}{\sqrt{B_{y}^2+B_{z}^2}}
\end{equation}
where $\alpha=\sqrt{\Gamma p/\rho}$ denotes the speed of sound. With this notation,
the right eigenvectors in matrix form will be given by
\begin{equation}
  \tens{R} = \left(\begin{array}{ccccccccc}
   0  &          \rho\alpha_f                 &             0                  &         \rho \alpha_s              & 1     &         \rho \alpha_s              &             0                  &  \rho\alpha_f                         & 0   \\ \noalign{\medskip}
   0  &          -c_f\alpha_f                 &             0                  &        -\alpha_s c_s               & 0     &         \alpha_s c_s               &             0                  &   c_f\alpha_f                         & 0   \\ \noalign{\medskip}
   0  &       \alpha_s c_s \beta_y S          & -\frac{\beta_z}{\sqrt{2}}      & -\alpha_f c_f \beta_y S            & 0     &  \alpha_f c_f \beta_y S            & -\frac{\beta_z}{\sqrt{2}}      &   -\alpha_s c_s \beta_y S             & 0   \\ \noalign{\medskip}
   0  &       \alpha_s c_s \beta_z S          &  \frac{\beta_y}{\sqrt{2}}      & -\alpha_f c_f \beta_z S            & 0     &  \alpha_f c_f \beta_z S            &  \frac{\beta_y}{\sqrt{2}}      &   -\alpha_s c_s \beta_z S             & 0   \\ \noalign{\medskip}
   1  &                0                      &             0                  &             0                      & 0     &             0                      &             0                  &           0                           & 1   \\ \noalign{\medskip}
   0  &\alpha_s \sqrt{\rho} \alpha \beta_y    & -\sqrt{\frac{\rho}{2}} \beta_z &-\alpha_f \sqrt{\rho} \alpha \beta_y& 0     &-\alpha_f \sqrt{\rho} \alpha \beta_y&  \sqrt{\frac{\rho}{2}} \beta_z &  \alpha_s \sqrt{\rho} \alpha \beta_y  & 0   \\ \noalign{\medskip}
   0  &\alpha_s \sqrt{\rho} \alpha \beta_z    &  \sqrt{\frac{\rho}{2}} \beta_y &-\alpha_f \sqrt{\rho} \alpha \beta_z& 0     &-\alpha_f \sqrt{\rho} \alpha \beta_z& -\sqrt{\frac{\rho}{2}} \beta_y &  \alpha_s \sqrt{\rho} \alpha \beta_z  & 0   \\ \noalign{\medskip}
   0  &           \alpha_f \Gamma p           &             0                  &        \alpha_s \Gamma p           & 0     &        \alpha_s \Gamma p           &             0                  &   \alpha_f \Gamma p                   & 0   \\ \noalign{\medskip}
 -c_h &                0                      &             0                  &             0                      & 0     &             0                      &             0                  &              0                        & c_h
\end{array}\right)\, 
\end{equation}
where $S=\rm{sign}(B_x)$. On the other hand, the left eigenvectors are
\begin{equation}
   \tens{L} = \left(\begin{array}{ccccccccc}
   0             &                0                         &             0                            &                 0                        &  \frac{1}{2}        &                 0                          &                 0                            &                0              &    -\frac{1}{2c_h}    \\ \noalign{\medskip}
   0             &-\frac{\alpha_f c_f}{2\alpha^2}           &\frac{\alpha_s c_s \beta_y S}{2\alpha^2}  &\frac{\alpha_s c_s \beta_z S}{2\alpha^2}  &       0             &\frac{\alpha_s\beta_y}{2\sqrt{\rho}\alpha}  & \frac{\alpha_s\beta_z}{2\sqrt{\rho}\alpha}   &\frac{\alpha_f}{2\rho\alpha^2} &            0          \\ \noalign{\medskip}
   0             &                0                         & -\frac{\beta_z}{\sqrt{2}}                &  \frac{\beta_y}{\sqrt{2}}                &       0             &    -\frac{\beta_z}{\sqrt{2\rho}}           &  \frac{\beta_y}{\sqrt{2\rho}}                &                0              &            0          \\ \noalign{\medskip}
   0             & -\frac{\alpha_s c_s}{2\alpha^2}          &-\frac{\alpha_f c_f \beta_y S}{2\alpha^2} &-\frac{\alpha_f c_f \beta_z S}{2\alpha^2} &       0             &-\frac{\alpha_f\beta_y}{2\sqrt{\rho}\alpha} &-\frac{\alpha_f\beta_z}{2\sqrt{\rho}\alpha}   &\frac{\alpha_s}{2\rho\alpha^2} &            0          \\ \noalign{\medskip}
   1             &                0                         &             0                            &                 0                        &       0             &                 0                          &             0                                &                 0             &   -\frac{1}{\alpha^2} \\ \noalign{\medskip}
   0             &  \frac{\alpha_s c_s}{2\alpha^2}          & \frac{\alpha_f c_f \beta_y S}{2\alpha^2} & \frac{\alpha_f c_f \beta_z S}{2\alpha^2} &       0             &-\frac{\alpha_f\beta_y}{2\sqrt{\rho}\alpha} &-\frac{\alpha_f\beta_z}{2\sqrt{\rho}\alpha}   &\frac{\alpha_s}{2\rho\alpha^2} &            0          \\ \noalign{\medskip}
   0             &                0                         & -\frac{\beta_z}{\sqrt{2}}                &  \frac{\beta_y}{\sqrt{2}}                &       0             &     \frac{\beta_z}{\sqrt{2\rho}}           & -\frac{\beta_y}{\sqrt{2\rho}}                &                0              &            0          \\ \noalign{\medskip}
   0             & \frac{\alpha_f c_f}{2\alpha^2}           &-\frac{\alpha_s c_s \beta_y S}{2\alpha^2} &-\frac{\alpha_s c_s \beta_z S}{2\alpha^2} &       0             &\frac{\alpha_s\beta_y}{2\sqrt{\rho}\alpha}  & \frac{\alpha_s\beta_z}{2\sqrt{\rho}\alpha}   &\frac{\alpha_f}{2\rho\alpha^2} &            0          \\ \noalign{\medskip}
   0             &                0                         &             0                            &                 0                        &  \frac{1}{2}        &                 0                          &                 0                            &                0              &     \frac{1}{2c_h}    
\end{array}\right).
\end{equation}

%
%



%
%



\end{document}